\documentclass[3p,times]{elsarticle}
\usepackage{ecrc}
\usepackage{amssymb}
\usepackage{amsmath}
\usepackage{booktabs}
\usepackage{multirow}
\usepackage{enumitem}
\usepackage{filecontents}
\usepackage[figuresright]{rotating}
\usepackage[bookmarks={false}, pdfstartview={XYZ null null 1.00}]{hyperref} 
\hypersetup{breaklinks=true, colorlinks= true, urlcolor= black, linkcolor= black, citecolor= black,filecolor= black}
\usepackage{lineno}
\usepackage{framed} 
\usepackage{multicol} 
\usepackage{nomencl} 
\volume{}

\firstpage{1}

\journalname{\normalsize \textup{International Journal of Multiphase Flow}}

\runauth{\normalsize \textit{Vankeswaram} et al.}

\jid{International Journal of Multiphase Flow}

\jnltitlelogo{\large International Journal of Multiphase Flow}

\CopyrightLine{2024}{Elsevier Ltd.}

\biboptions{authoryear}

\renewcommand*\nompreamble{\begin{multicols}{2}}
\renewcommand*\nompostamble{\end{multicols}}

\usepackage{etoolbox}
\renewcommand\nomgroup[1]{%
  \item[\bfseries
  \ifstrequal{#1}{A}{Latin Symbols}{%
  \ifstrequal{#1}{B}{Greek Symbols}{%
  \ifstrequal{#1}{C}{Acronyms}{}}}%
]}
\makenomenclature
\begin{document}

\begin{frontmatter}



\dochead{Manuscript Preprint}

\title{Spatial evolution of droplet size and velocity \\ characteristics in a swirl spray}


\author[1]{S. K. Vankeswaram\corref{cor1}\fnref{fn1}}
\ead{krishnavankeswaram@gmail.com}
\cortext[cor1]{Corresponding author}
\fntext[fn1]{Currently at Brunel University London, Uxbridge UB8 3PH, United Kingdom}
\author[2]{V. Kulkarni}
\author[1]{S. Deivandren}
\address[1]{Department of Aerospace Engineering, Indian Institute of Science, Bangalore 560012, India}
\address[2]{Harvard University, Cambridge, MA 02138, USA}

\begin{abstract}
Spray drop size distribution generated by atomization of fuel \textcolor{black}{influences} several facets of a combustion process such as, fuel-air mixing, reaction kinetics and thrust generation. In a typical spray, the drop size distribution evolves spatially, varying significantly between the near and far regions of the spray. However, studies so far have focused exclusively on either one of these regions and are unclear on the exact axial location where transition from near to far region droplet size characteristics is expected. In this work, we address this crucial gap by considering a swirl atomizer assembly and measuring the droplet characteristics for different liquid flow conditions of the ensuing spray at various radial and axial locations. Our results reveal an undiscovered axial variation in the scaled radial droplet velocity profiles, not followed by \textcolor{black}{the} radial drop size profiles, from which we unambiguously demarcate the near region as the zone which extends up to axial distances of 2.0 to 2.5 times film breakup length. Beyond this distance, the drop size characteristics are influenced by external factors such as airflow and identified as the far region of the spray. \textcolor{black}{Using our} analysis we locate the point of origin of the commonly reported droplet high-velocity stream along the spray centreline to the end of film breakup or near region of the spray. We also \textcolor{black}{find that} the global probability density functions for droplet size and velocity which show a marked difference in the near and far regions; being bimodal in the near-region and unimodal in the far-region being well represented by the double Gaussian and the Gamma distributions, respectively. \textcolor{black}{We further quantify} our results by meticulous measurements of number and volume flux distributions, global mean drop sizes, drop size ($D_d$) axial velocity ($U_a$) correlations, axial velocity based on drop size classification and turbulent kinetic energy (TKE) which reveal the effect of drop inertia and air flow in \textcolor{black}{determining the} statistics in both the near and far regions. We anticipate the findings of this work \textcolor{black}{will} guide \textcolor{black}{future} investigations on combustion processes and combustor design based on spray characteristics.
\end{abstract}


\end{frontmatter}
\clearpage

\begin{table*}[!t]   
\begin{framed}
\nomenclature[A]{\textit{Re}}{Reynolds number of the liquid}
\nomenclature[A]{\textit{We}}{Weber number of the liquid}
\nomenclature[A]{$U_l$}{liquid sheet streamwise velocity\dotfill m/s}
\nomenclature[A]{$U_a$}{droplet axial velocity\dotfill m/s}
\nomenclature[A]{$U_r$}{droplet radial velocity\dotfill m/s}
\nomenclature[A]{$U_y$}{droplet tangential velocity\dotfill m/s}
\nomenclature[A]{$D_d$}{drop size\dotfill $\mu\textrm{m}$}
\nomenclature[A]{$P_b$}{gamma distribution probability function}
\nomenclature[A]{$r$}{radial (horizontal) coordinate\dotfill mm}
\nomenclature[A]{$r^{*}$}{dimensionless radial coordinate}
\nomenclature[A]{$y$}{tangential (in-plane) coordinate\dotfill mm}
\nomenclature[A]{$z$}{axial coordinate\dotfill mm}
\nomenclature[A]{$z^{*}$}{dimensionless axial coordinate}
\nomenclature[A]{$u^{\prime}$}{axial fluctuating droplet velocity\dotfill m/s}
\nomenclature[A]{$v^{\prime}$}{radial fluctuating droplet velocity\dotfill m/s}
\nomenclature[A]{$w^{\prime}$}{tangential fluctuating droplet velocity\dotfill m/s}
\nomenclature[A]{\textit{t}\textsubscript{\textit{f}}}{thickness of conical sheet\dotfill $\mu\textrm{m}$}
\nomenclature[A]{$n$}{number flux}
\nomenclature[A]{$n_{\textrm{\textit{N}}}$}{normalized number flux}
\nomenclature[A]{$q_{\textrm{\textit{N}}}$}{normalized volume flux}
\nomenclature[A]{$t_{\textrm{\textit{aq}}}$}{data acquisition time\dotfill s}
\nomenclature[A]{$Z_{\textrm{\textit{NR}}}$}{length of near region from orifice exit\dotfill mm}
\nomenclature[A]{$Z_{\textrm{\textit{b}}}$}{length of film breakup from orifice exit\dotfill mm}
\nomenclature[B]{$\alpha$}{gamma distribution parameter}
\nomenclature[B]{$\varphi$}{log-normal distribution function}
\nomenclature[B]{$\beta$}{spray cone angle\dotfill degrees}
\nomenclature[B]{$\rho_l$}{liquid density\dotfill kg/m\textsuperscript{3}}
\nomenclature[B]{$\mu_l$}{liquid viscosity\dotfill mPa-s}
\nomenclature[B]{$\sigma_l$}{liquid surface tension\dotfill mN/m}
\nomenclature[B]{$\Gamma$}{gamma function}
\nomenclature[B]{$\zeta$}{similarity variable}
\nomenclature[C]{SMD}{sauter mean diameter}
\nomenclature[C]{PDF}{probability distribution function}
\nomenclature[C]{PDI}{phase doppler interferometry}
\nomenclature[C]{TKE}{turbulent kinetic energy}
\nomenclature[C]{RMS}{root mean squared}
\printnomenclature
\end{framed}
\end{table*}

\section{Introduction}\label{Sec1}
The process of atomization, which is the transformation of bulk fluid into tiny droplets is routinely observed in various industrial applications such as spray combustion, food processing, pharmaceutical sprays, spray painting, agricultural sprays, and naturally occurring phenomena such as mist formation and rain fall \citep{Sivakumar2016, Lefebvre2017,Vankeswaram2023,Fong2024}. In gas turbine fuel combustion, the role of atomization is to increase the specific surface area of the fuel to provide better mixing and evaporation for efficient combustion \citep{Lefebvre2017}. Atomization typically materializes as a two-step process beginning with the primary breakup of liquid sheet/jet of the fuel into the small droplets culminating with secondary atomization \citep{Kulkarni2014}, where these smaller droplets undergo further disintegration \citep{Lefebvre2017}. Expectedly, both primary and secondary atomization are important for controlling the fuel-air mixing rates in a gas turbine \textcolor{black}{combustors} for a spray atomizer \citep{An2023, Faeth1983} and rely heavily on atomizer geometry\citep{Nikouei2023, Rajesh2023}. While average droplet size and velocity characteristics convey crucial information about these sprays, the spatial evolution of spray characteristics like droplet size and velocity can differ greatly from these average values as the spray evolves from near the atomizer exit to regions far away from it. Thus, a detailed understanding of their spatial evolution can contribute tremendously towards improved fuel-air mixing and better droplet combustion in gas \textcolor{black}{turbine combustors} where primarily average values have been used \citep{Faeth1983, Jones2014}. In other applications such as agricultural pesticide sprays where spray drift must be controlled and, paint, pharmaceutical and food processing sprays, where a uniform coating must be applied, information on spatial spray characteristics can help process optimization which has largely been based on average (drop size and velocity) values.

In an effort to address the above, research so far has identified three distinct zones of the spatial spray morphology with differing spray attributes\citep{Saha2012}. First of these is the liquid film to ligament production zone, where primary breakup takes place called the \textit{near region} of the spray. It is followed by secondary breakup zone, wherein the ligaments breakup into smaller droplets or the \textit{intermediate region} and, lastly, we have the third zone called the \textit{far region} where individual droplets produced in the previous zone coalescence to form larger droplets or breakup into smaller ones. Against this background, the transition from the first to the second zone is the most important since it delineates the near and far regions of the spray which have been known for vastly differing spray characteristics. Surprisingly, heretofore, these regions have only been distinguished approximately, by criteria such as distance from the nozzle, droplet sizes, spray cone angle or velocity \citep{Heindel2018}. A more systematic approach would rely on identifying key features of these near and far regions guiding efforts in delimiting these zones more precisely. To this end, we review the literature on spray characteristics in both near and far regions to highlight their salient attributes and contextualize how they may be used in further investigations. Though atomizer geometry and injections strategy is an important consideration in analyzing any spray morphology we limit our attention to pressure swirl sprays. Our choice is motivated by the fact that conical sprays produced by swirl atomization offer advantages of finer atomization leading to high energy efficiency \citep{Lefebvre2000, Dafsari2017,Joseph2020}. 

The near region of the spray, as mentioned above, is approximately located close to the atomizer exit and constitutes the primary breakup zone where the liquid film breaks up \citep{Vankeswaram2022,Vankeswaram2024}. The apsherical ligaments produced by this breakup breakdown further to form large drops \citep{Schmidt1999, Lefebvre2017}. The relatively large size of these droplets renders them unstable and are further destabilized by the higher velocities they possess due to their proximity to the injector exit. Imaginably, these drops display lower residence times, higher levels of turbulence intensities and number densities. Last of these, \textit{i}.\textit{e}. the higher number density leads to higher droplet collisions \citep{Pawar2015} which influences the overall drop size distribution statistics. Furthermore, pressure swirl atomization patterns in this region can also have an effect of the droplet sizes produced. In this regard, Santolaya et al.\citep{Santolaya2007} have shown that as the breakup transitions from perforated sheet to wavy sheet \citep{Sivakumar2011} a finer spray with a higher radial dispersion is obtained, which also shows increased axial velocities at the core due to higher air entrainment. The liquid sheet and the droplets transfer momentum to the air \citep{Jedelsky2018} which forms the core of the spray. \textcolor{black}{At low injection pressures this has been reported to lead to an intriguing behavior in the radial drop size and velocity profiles which is also seen for spray characteristics in the far region as discussed below.}

At large distances from the injector exit, in the far region of the spray, the droplets produced undergo substantial reduction in size due to secondary breakup and/or evaporation. As a consequence, the spray in this region is less turbulent, slower in speed possessing longer residence times. The near spherical nature of these drops and the absence of aspherical ligaments makes this zone amenable for experimental drop size-velocity measurements using several experimental techniques and has unsurprisingly been the focus of many investigations \citep{Saha2012, Durdina2014, Sivakumar2015, Dhivyaraja2019, Vankeswaram2022}. Recent studies \citep{Brenn2016, Hinterbichler2020} have unearthed an interesting  self-similar behavior for the drop(liquid) sizes and gas phase velocity using the smallest spray droplets as tracer particles from PDI data. The implications of such findings are profound as self-similarity implies a reduction in dependency of spray characteristics such as drop size and velocity on multiple geometric and kinematic variables. This work has been extended to establish dynamic similarity in swirl sprays for different atomizers by Dhivyaraja et al.\citep{Dhivyaraja2019}.Their work also showed that self-similarity using specific geometric scales can be used to identify the core from the periphery of the spray where drop ballistics become important. Follow up studies have shown self-similar behavior in pressure swirl sprays of different configurations such as those with recessed orifices \citep{Joseph2020}, airblast \citep{Soni2021} and inner air swirl \citep{Li1999}. In the near region too, when the effect of recirculating air flow is small as seen at low injection pressures self-similar behavior of droplet mean axial velocity and Sauter Mean Diameter (SMD) for a hollow cone swirl spray has been reported \citep{Vankeswaram2022}. In the simple analysis employed in many of these studies, liquid sheet thickness ($t_f$) for drop sizes and the distance at which axial velocity decays to a certain percentage of its value at the centerline for the radial direction, is used to obtain the self-similar curves.

\textcolor{black}{So far in our discussion we have focused our attention on cold flow, non-reacting conditions at ambient temperature and pressure. However, at engine relevant, reactive flow conditions, several other factors such as, elevated temperature and pressure, physical properties of the fuel could be important. Especially since heat release during combustion and chemical reactions can modify the physical properties of the fuels  \hypersetup{citecolor=black}{\citep{Turns1996}}. The formation of new products during combustion can also influence interactions between spray and ignition front affecting flame structure and stability  \hypersetup{citecolor=black}{\citep{Hodge2024, Philo2023}}. Smaller drops combust faster, increasing efficiency of combustion and in turn increasing local temperature and evaporation rates  \hypersetup{citecolor=black}{\citep{Chigier1974}}.}

\textcolor{black}{Prior studies  \hypersetup{citecolor=black}{\citep{Kannaiyan2020, Kannaiyan2021, Kannaiyan2023, Kannaiyan2024}} have investigated these characteristics in detail for different such as Gas-to-Liquid (GTL) jet fuel, conventional Jet A-1 and NG-SPK (Natural Gas-based Synthetic Paraffinic Kerosene) jet fuel at varying ambient gas temperatures and pressures. Overall, microscopic drop sizes and macroscopic features such as spray cone angle, liquid sheet breakup length were found to be dependent on these factors and varying spatially. For instance, it is reported that the sheet breakup region for the GTL jet fuel is slightly longer and spray width wider (in far regions) than that of the conventional Jet A-1 fuel  \hypersetup{citecolor=black}{\citep{Kannaiyan2020}} at elevated ambient temperatures and pressure. Similarly, it was documented that the droplet mean axial velocities are influenced by ambient gas temperature dependent fuel properties close to the nozzle exit  \hypersetup{citecolor=black}{\citep{Kannaiyan2021}} and overall, for GTL jet fuels smaller droplet and Sauter mean diameters  \hypersetup{citecolor=black}{\citep{Kannaiyan2023}} compared to conventional fuels. Between fuels, NG-SPK jet fuel was shown to have a higher percentage of smaller droplet diameters than Jet A-1 fuel  \hypersetup{citecolor=black}{\citep{Kannaiyan2024}}. These studies show that indeed spray characteristics of fuels can evolve spatially at engine relevant conditions and effect which can possibly be seen in cold flow conditions and help us underscore the significance of these earlier works.}

From the foregoing review, we identify the following gaps in existing literature which we aim to fill through this work. To begin, we differentiate the near and far regions by axial location, quantitatively from the drop size and velocity data obtained from PDI (Phase Doppler Interferometry). The premise for this is adopted from the work of Dhivyaraja et al.\citep{Dhivyaraja2019} who used scaled radial variation of drop size for identifying the core of the spray. We extend their work, albeit with appropriate choice of length and velocity scales to demarcate axially the near and far regions precisely. Based on this, two axial (\textit{Z}) locations are identified, one in the near-region, close to the breakup of the liquid sheet ($Z = Z_b$) and the other far away ($Z = 30$ mm $\approx 2.5 Z_b$) from the liquid film breakup for further detailed analysis. Several unknown aspects are unearthed, such as, it is unknown whether gamma or log-normal distribution best describes the global drop size distributions. We also plot the global (for all radial locations at a given axial location) drop size ($D_d$)-velocity($U_a$) correlations to observe any discernible difference in the near and the far region. It is known that far away from the nozzle $D_d - U_a$ correlation shows two peaks, whether this followed in the near region is not known. Also, undocumented is the axial velocity variation for drops of different sizes as distributed in bins of different drop size classes. Since, smaller drops follow the air velocity they are representative of the air flow field. In contrast, large ones have appreciable inertia and behave differently. How this would vary from the near to far region is unknown and is explored in this work. Lastly, we analyze the turbulence intensity in these sprays and how they evolve spatially.

In the view of the objectives outlined in the previous paragraph, the paper is organized along the following lines, Section \ref{Sec1} describes the context, background and motivation for this study. In Section \ref{Sec2} experimental details and test conditions are provided. Details of our results (Section \ref{SecR}) begin with Section \ref{Sec3} wherein we identify appropriate scales to help demarcate near and far regions of the spray based on radial variation of drop sizes and axial velocity with \textit{We}. Following this, in Section \ref{Sec4} contour plots of normalized resultant total velocity, $U_{\textrm{\textit{res}}}/U_l$ and drop sizes are shown to understand the entire spray evolution. The discussion thereafter is based on providing details at two select locations in the near ($Z = Z_b$) and far ($Z =$ 30 mm) region of the spray. To begin, in Section \ref{Sec5} we describe the number and volume flux distributions in these regions followed by a description of the drop size and velocity correlation in Section \ref{Sec6}. After this, the mean drop size characteristics elucidated in Section \ref{Sec7}. Since the entrainment of drops caused by the air flow usually redistributes them spatially, we expand on our data on mean drop sizes by examining their probability distributions locally and globally in Section \ref{Sec8}. Moving ahead we classify the drops sizes into various class sizes and plot axial velocity characteristics in near and far region in Section \ref{Sec9}. The documentation of our results ends with the characterization of turbulent intensity in the spray in Section \ref{Sec10} and in closing we summarize our results and provide our final remarks. 

\section{Experimental Methods and Test Conditions}\label{Sec2}
\begin{figure*}[t!]
   \centering
   \vspace{0pt}
    \includegraphics[width=\textwidth]{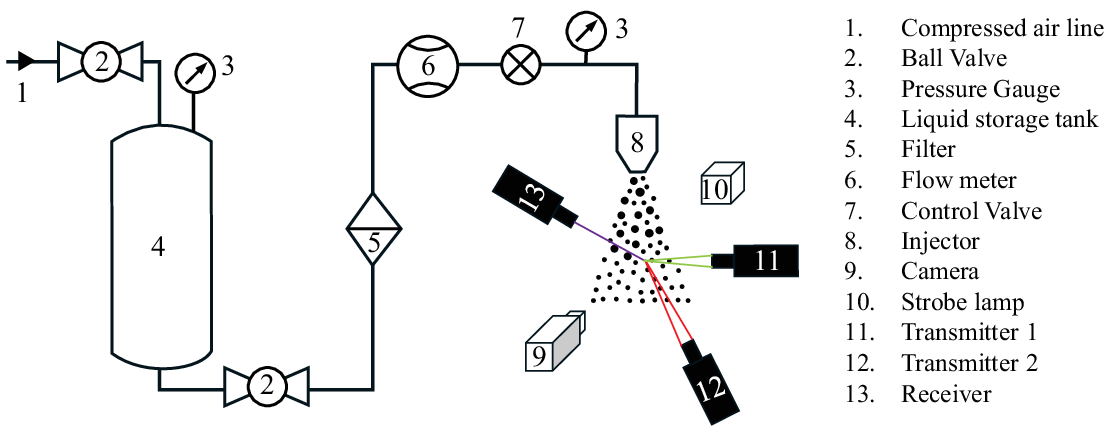}
   \vspace{-15pt}
  \caption{\label{Fig1}Schematic of spray test facility.}
\end{figure*}

Schematic of the spray test facility used to collect the experimental measurements for water sprays discharging from a hollow cone swirl nozzle is shown in Fig. \ref{Fig1}. The physical properties of water are density, $\rho_l$ = 997 kg/m\textsuperscript{3}, viscosity, $\mu_l$ = 1.00 $\times$ 10\textsuperscript{-3} Ns/m\textsuperscript{2}, and surface tension, $\sigma_l$ = 0.072 N/m at temperature 20 $^{\circ}$C) are taken from Kundu and Cohen \citep{Kundu2015}. The facility is equipped with the required flow control elements such as the ball valves (2), filter (5), control valves (7), pressure gauges (3), and flow meter (6) and is connected to the injector flow supply line. The mass flow rate of the liquid flow to the injector is provided by the flow meter (Micro Motion\textsuperscript{\textregistered} R-Series Coriolis type flow meter, Emerson) with an accuracy of 0.75\% of rate. A detailed description of the experimental setup is provided in our previous work \citep{Vankeswaram2022}.

The spray images for different flow conditions are captured with a spatial resolution of 31.3 $\mu\textrm{m}$ per pixel using a digital camera (9) (Nikon\textsuperscript{\textregistered} D7100 DSLR camera, Nikon\textsuperscript{\textregistered}) equipped with a zoom lens (AF Zoom Nikon\textsuperscript{\textregistered} 80-200mm f/2.8D, Nikon\textsuperscript{\textregistered}). The drop characteristics such as the mean drop size and velocity components are measured using a 3D PDI (3 Dimensional Phase Doppler Interferometry) system (manufactured by Artium Technologies) as shown in Fig. \ref{Fig1}. The system is comprised of two transmitters (11 and 12), one receiver (13), an advanced signal analyzer unit, and a computer-controlled three-dimensional traverse unit (not explicitly shown here). The focal length of the laser transmitter and receiver lenses is kept at 750 mm and 530 mm respectively, for all experiments. The transmitters and receiver are positioned on the traverse system in such a way that the receiver is 30$^{\circ}$ off-axis with respect to the transmitters in forward scatter mode \textcolor{black}{such that maximum intensity of the scattered first order refraction from the droplet is captured}. Further details on the operating procedure of the equipment can be found in \citep{Bachalo1984,Albrecht2013,Joseph2020} and the measurement details in \citep{Vankeswaram2022}.
\begin{figure*}[b!]
   \centering
   \vspace{0pt}
    \includegraphics[scale = 0.8]{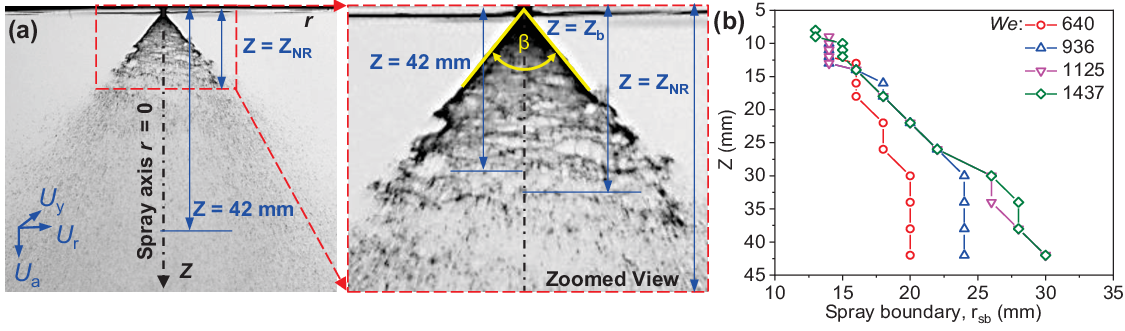}
   \vspace{-15pt}
  \caption{\label{Fig2}(\textit{a}) Water spray with $We = 1437$ from the swirl atomizer. (\textit{b}) Variation of $r_{sb}$ with downstream axial location, $Z$ for sprays with different \textit{We} from the swirl atomizer.}
\end{figure*}
Information on the operating conditions and flow parameters is presented in Table \ref{Table1}. The details on the estimation of liquid sheet streamwise ($U_l$) and axial ($U_{ax}$) velocity are provided in \citep{Joseph2020,Vankeswaram2022}. Also, in Table \ref{Table1} the liquid Weber number, \textit{We} is given by, $\rho_lU_l^2t_f/\sigma_l$ and the Reynolds number, \textit{Re} by $\rho_lU_lD_0/\mu_l$ where, $t_f$ is liquid film thickness measured using the calculations detailed previously \citep{Kulkarni2010} and $D_0 \approx 0.69$ mm is the exit orifice scale for the liquid sheet. The mass flow rate, $\dot{m}$ for a given liquid pressure difference ($\Delta P$) is directly measured using the flowmeter (shown schematically in Fig. \ref{Fig1}).
 
Fig. \ref{Fig2}\textcolor{black}{(a)} shows a representative image of water spray discharging from the swirl atomizer with $\dot{m} =  $4.55 \textit{g}/\textit{s}, $We =$1437 and $Re =$ 18042. Various spray parameters of significance, such as the spray cone angle, $\beta$, liquid film breakup distance ($L_b$), the location of near-region PDI measurement location ($Z_b$), the axial location demarcating the near and far region of the spray ($Z_{NR}$), and the maximum axial location ($Z =$ 42 mm) at which the drop size measurements are made are marked in Fig. \ref{Fig2}\textcolor{black}{(a)}. Radial coordinate as used to describe PDI data, strictly speaking is the Cartesian horizontal coordinate, however we use the term ``\textit{radial}'' for simplicity. Similarly, the term ``\textit{tangential}'' is used for in-plane coordinate shown as ``\textit{y}'' in Fig. \ref{Fig2}\textcolor{black}{(a)}. Detailed information on the variation of breakup location, $L_b$ with \textit{We} for conical swirl sprays are in given prior studies \citep{Kulkarni2010, Sivakumar2011}. In this study, $L_b$ is experimentally measured from the spray images captured for different \textit{We} using backlit shadowgraphs. \textcolor{black}{From a given image (shown in Fig. \hypersetup{linkcolor=black}{\ref{Fig2}}\textcolor{black}{(a)} with a zoomed view to the right), $L_b$ is seen to be non-uniform circumferentially and taken as the average of maximum and minimum values of film breakup length. This is repeated for a set of 45 images to obtain the final value. On the other hand, for identifying $Z_b$ we first calculate the standard deviation of $L_b$ and make PDI measurements at locations in the vicinity of the $L_b \pm$ standard deviation. The locations where the validation rates are well above the critical values (above 60\%) are then used as the threshold for acceptability of measurements and identification of $Z_b$. Since the value of $Z_b$ rests on the value of $L_b$ its uncertainty may be assumed to be the same as that of $L_b$ and less than 5.6\%.}

\begin{table}[htp!]
\small
\begin{center}
{ 
\caption{\normalsize Injection parameters and flow characteristics.}\label{Table1}
{\begin{tabular}{cccccccc}
\\
\toprule
  $\Delta P$ & \hspace{0.3em} $\dot{m}$ & \hspace{0.3em} ${\beta}$ & \hspace{0.3em} $U_l$ & \hspace{0.3em} $U_{ax}$ & \hspace{0.3em} $t_f$ & \hspace{0.3em} $We\;$ & \hspace{0.3em} $Re\;$\\
  (MPa)  &\hspace{0.3em} (\textit{g/s}) &\hspace{0.3em} (\textit{deg}) & \hspace{0.3em}$(m/s)$ & \hspace{0.3em}$(m/s)$ & \hspace{0.3em}$(\mu\textrm{m})$ & (-) & (-)\\
\midrule
     0.97  & \hspace{0.3em} 3.21 & \hspace{0.3em} 74.0 & \hspace{0.3em} 21.3 & \hspace{0.3em} 17.0 & \hspace{0.3em} 102.4 & 640 & 11706\\
     \\[-1em]
     1.38 & \hspace{0.3em} 3.76 & \hspace{0.3em} 76.4  & \hspace{0.3em} 26.3  & \hspace{0.3em} 20.7  & \hspace{0.3em} 97.9 & 936 & 14254 \\
      1.65 & \hspace{0.3em} 4.09 & \hspace{0.3em} 76.5  & \hspace{0.3em} 29.2 & \hspace{0.3em} 22.9  & \hspace{0.3em} 95.7 & 1125 & 15769 \\
     2.07 & \hspace{0.3em} 4.55& \hspace{0.3em} 77.3  & \hspace{0.3em} 33.5 & \hspace{0.3em} 26.2 & \hspace{0.3em} 92.8 & 1437 & 18042 \\
\bottomrule
\end{tabular}
    }}
\end{center} 
\end{table}

As will be seen in later discussion, identification of spray boundary (denoted by subscript \textit{sb} hereafter) location, $r_{sb}$ at a given $Z$ in the spray is essential for the analysis. Fig. \ref{Fig2}\textcolor{black}{(b)} shows the variation of $r_{sb}$ with $Z$ at different \textit{We} for the spray from the swirl atomizer. The spray boundary, $r_{sb}$ is chosen at a given $Z$ as the radial location at which the droplet number count is 5\% of the maximum droplet number count at that $Z$ location \citep{Hinterbichler2020}. In the present case of atomizer, the spray flow transitions from diverged tulip- shaped spray to triangular-shaped conical spray at \textit{We} between 640 to 936. It can be seen from Fig. \ref{Fig2}\textcolor{black}{(b)} that the spray diverges only up to a certain $r$ for 640 to 936, but after that the value of $r_{sb}$ (spray boundary) remains constant until the maximum $Z$ at which the measurements, are made in the present case ($Z$ = 42 mm). 

For sprays with $We =$ 1125 and 1437, the spray boundary continuously diverges with $Z$. At $Z$ close to the liquid sheet breakup ($Z$ $\sim$ 10 - 22 mm), not much difference in $r_{sb}$ is seen with \textit{We}, which is in line with observations reported in previous studies \citep{Durdina2014}. But at $Z$ far from the liquid sheet breakup ($Z >$ 22 mm), the liquid spray diverges with increase in \textit{We}, due to an increase in the kinetic energy of the liquid sheet with increasing \textit{We} which promotes the radial dispersion of the atomized droplets for longer axial distances ( \hypersetup{citecolor=black}{\cite{Liu2019}}; \cite{Durdina2014}). \textcolor{black}{Lastly, the PDI ch1 validation rates are well above 60\% in the near-region and above 85\% in the far-region of film breakup confirming the high sphericity of the droplets in those regions and underscoring the effectiveness of our experimental technique  \hypersetup{citecolor=black}{\citep{Vankeswaram2022}}.}

\section{Results and Discussion}\label{SecR}
\subsection{\textbf{Selection of scales and demarcation of near and far regions of the spray}}\label{Sec3}
Different studies used different scaling parameters to normalize the mean droplet size and mean velocity \hypersetup{citecolor=black}\citep{Li1999,Hinterbichler2020,Joseph2020,Soni2021,Vankeswaram2022}. It was identified by the authors in their earlier work carried out for the same injector that the spray at the radial location corresponding to maximum droplet volume flux is the representative of the global spray in the near region of liquid film breakup\citep{Vankeswaram2022}. The choice of an alternate similarity variable used in previous work \citep{Dhivyaraja2019} is evaluated in \ref{SecAppA} and found to be unsuitable for present analysis. 
\begin{figure*}[htp!]
   \centering
   \vspace{0pt}
    \includegraphics[scale=1.0]{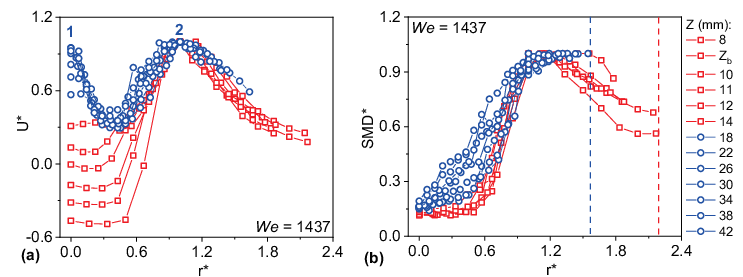}
   \vspace{-10pt}
  \caption{\label{Fig3}Variation of scaled (a) droplet mean velocity, and (b) SMD along the scaled radial distance at different axial locations for $We =$ 1437. In (a) the peak of the velocity distributions are marked by \textcolor{black}{\textbf{1}} and \textcolor{black}{\textbf{2}} while in (b) the dotted lines represent the edge of the spray.}
\end{figure*}
So, in the present work, the droplet axial velocity ($U_a$) and SMD (Sauter Mean Diameter) at a given \textcolor{black}{spatial location} ($Z$, $r$) is normalized using the maximum axial velocity and maximum SMD at the corresponding ($Z, r$) and the radial distance, $r$ is normalized by the radial distance that corresponds to the maximum volume flux location, $r_{mvf}$ (also see Section \ref{Sec5}) and are expressed as,
\begin{equation}\label{simvaw}
U^{*} = \dfrac{U_a\left(Z, r\right)}{U_{max}\left(Z, r\right)}, \quad\quad {\textrm{SMD}}^{*} = \dfrac{\textrm{SMD}\left(Z, r\right)}{\textrm{SMD}_{max}\left(Z, r\right)}, \quad\quad r^{*} = \dfrac{r}{r_{mvf}}
\end{equation}
Using Eq. \ref{simvaw}, Fig. \ref{Fig3} shows the variation of normalized droplet axial velocity ($U^{*}$), and normalized SMD (${\textrm{SMD}}^{*}$) with the normalized radial distance ($r^*$) at different axial locations ($Z$) for $We =$ 1437. \textcolor{black}{Through two trials, the measurement uncertainty involved for dimensional SMD and U in these measurements are $\pm$2.5\% and $\pm$3\% respectively}. It can be seen from Fig. \ref{Fig3}\textcolor{black}{(a)} and \textcolor{black}{(b)} that both $U^{*}$ and ${\textrm{SMD}}^{*}$ show two distinct variations with $r^{*}$ at different $Z$. Until $Z =$ 14 mm (marked by red square symbols in the figure), both $U^{*}$ and ${\textrm{SMD}}^{*}$ show a similar variation: the value of $U^{*}$ and ${\textrm{SMD}}^{*}$ almost remains constant up to certain $r^{*}$, thereafter sharply increases to a maximum value with increase in $r^{*}$, and then decreases towards the spray periphery. The initial constant value of $U^{*}$ and ${\textrm{SMD}}^{*}$ is attributed to the presence of recirculatory vortex in the spray, as it can be seen from Fig. \ref{Fig3}\textcolor{black}{(a)} that $U^{*}$ is changing sign from negative to positive at $r^{*}$ near the spray axis ($r^{*} = 0$). With the increase in $r^{*}$, $U^{*}$ and ${\textrm{SMD}}^{*}$ increase as the measurement location approaches towards the liquid film breakup zone (spray production zone from the liquid film), and the values $U^{*}$ and ${\textrm{SMD}}^{*}$ decrease with $r^{*}$ moving away from the spray production zone. 

Thus, the variation of $U^{*}$ and ${\textrm{SMD}}^{*}$ with $r^{*}$ at $Z \le 14$ mm can be attributed to the behavior of liquid film breakup and therefore is influenced by the liquid film breakup \citep{Jedelsky2018}. For \textit{Z} $>$ 14 mm, the behaviour of $U^{*}$ and ${\textrm{SMD}}^{*}$ is different compared to \textit{Z} $\le$ 14 mm. It can be seen from Fig. \ref{Fig3}\textcolor{black}{(a)} that $U^{*}$ shows two peak values (marked by blue colored numbers \textcolor{black}{\textbf{1}} and \textcolor{black}{\textbf{2}}), one at the spray axis and the other near the spray periphery. The value of $U^{*}$ initially decreases with $r^{*}$, reaches a minimum, and increases thereafter until it reaches a maximum value. The maximum value of $U^{*}$ observed at the spray axis is reported in several early works \citep{Santolaya2007,Durdina2014,Du2017,Jedelsky2018}. From the point of the second maximum (point 2), the value of $U^{*}$ again decreases with the increase in $r^{*}$ towards the spray periphery. 
\begin{figure*}[t!]
   \centering
   \vspace{0pt}
    \includegraphics[width=\textwidth]{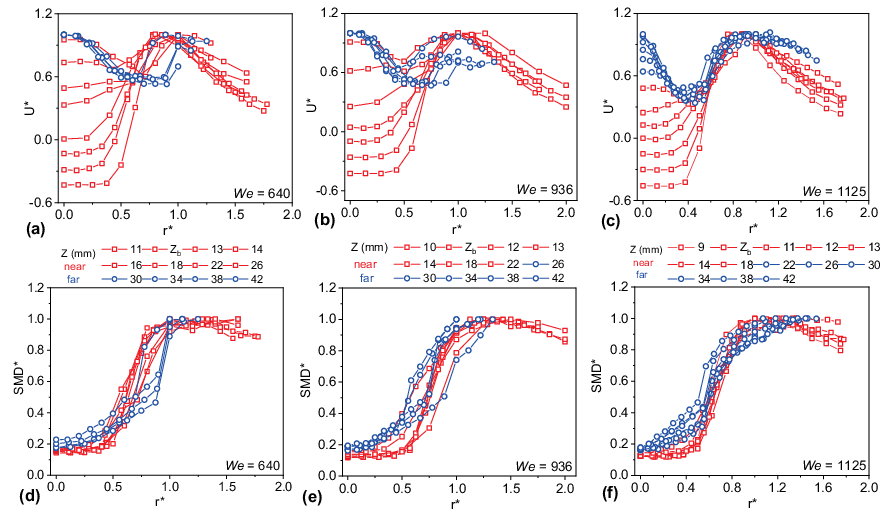}
   \vspace{-15pt}
  \caption{\label{Fig4}Variation of $U^{*}$ along $r^{*}$ at different $Z$ for the spray emerging from the swirl atomizer with different \textit{We} (a) 640, (b) 936, and (c) 1125. Similar variation for $\textrm{SMD}^{*}$ at \textit{We} (d) 640, (e) 936, and (f) 1125}
\end{figure*}
The droplets at the spray boundary are affected by the drag exerted by surrounding airflow, which decreases the droplet velocity at the spray boundary \citep{Saha2012, Durdina2014, Du2017} (spray boundary is highlighted by blue dotted line on Fig. \ref{Fig3}\textcolor{black}{(b)}  for the far-region of liquid film breakup). The value of  ${\textrm{SMD}}^{*}$ steadily increases with $r^{*}$, reaching peak at the spray periphery. Larger droplets are carried towards the spray periphery due to the centrifugal effect \citep{Saha2012, Durdina2014} which could be the reason for the maximum value of SMD at the spray periphery. Thus, for $Z >$ 14 mm, the spray characteristics are influenced by the external factors such as surrounding airflow. But it can also be observed from Figs. \ref{Fig3}\textcolor{black}{(a)} and \textcolor{black}{(b)}  that there are some axial locations ($Z =$ 14 mm and 18 mm) which are influenced by both the liquid film breakup and the external factors.

\begin{table}[htp!]
\small
\begin{center}
{ 
\caption{\normalsize Details of $Z_{NR}$ for the spray with different \textit{We} from the swirl atomizer.}\label{Table2}
{\begin{tabular}{cccc}
\\
\toprule
  \textit{We} & \hspace{0.3em} $L_b$ ($mm$) & \hspace{0.3em} $Z_b$ ($mm$) & \hspace{0.3em} $Z_{NR}$ ($mm$) \\
\midrule
     640 & \hspace{0.3em} 9.9 $\pm$ 0.52 & \hspace{0.3em} 12 & \hspace{0.3em} 24 (22-26)\\
     \\[-1em]
     936 & \hspace{0.3em} 8.7 $\pm$ 0.48 & \hspace{0.3em} 11 & \hspace{0.3em} 20 (18-22)\\
     \\[-1em]
     1125 & \hspace{0.3em} 8.1 $\pm$ 0.53 & \hspace{0.3em} 10 & \hspace{0.3em} $\sim$ 14-18\\
     \\[-1em]
     1437 & \hspace{0.3em} 7.2 $\pm$ 0.66 & \hspace{0.3em} 9 & \hspace{0.3em} $\sim$ 14-18\\
     \\[-1em]
\bottomrule
\end{tabular}
    }}
\end{center} 
\end{table}

From the above analysis, it is seen that the drop size and velocity along the axial and radial direction exhibits two distinct variations. Based on the spray droplet characteristics, the spray field can be divided into two regions, demarcated at $Z = Z_{NR}$ from the  atomizer exit. (1) The near-region of the liquid film breakup ($Z < Z_{NR}$), where the spray droplet characteristics are influenced primarily by the film breakup behavior, and (2) the far-region of the liquid film breakup ($Z > Z_{NR}$), where the spray droplet characteristics are influenced mainly by the external factors and the droplet inertia. It should be mentioned that, for a given spray flow condition, the value of $Z_{NR}$ is found to be within a range of $Z$, where the spray droplet characteristics are influenced by both the film breakup behavior and external factors. 

The variation of $U^{*}$ (see Fig. \ref{Fig4}\textcolor{black}{(a)}-\textcolor{black}{(c)}) and ${\textrm{SMD}}^{*}$ (see Fig. \ref{Fig4}\textcolor{black}{(d)}-\textcolor{black}{(f)}) with $r^{*}$ at different \textit{Z} for the spray with different \textit{We} from the swirl atomizer. It can be observed that the variation of $U^{*}$ and ${\textrm{SMD}}^{*}$ with $r^{*}$ show two distinct variations, like that observed with $We =$ 1437, at all \textit{We}. The scaling given in Eq. \ref{simvaw} demarcates the two regions well. For the spray with $We =$ 936 $-$ 1437, $U^{*}$ scaled very well at all $r^{*}$ in the far-region of liquid film breakup. Using this analysis, the demarcation of near and far region of the liquid film breakup is carried out and the values of $Z_{NR}$ at different \textit{We} so obtained are listed in Table \ref{Table2}.

\subsection{\textbf{Contour plots of resultant velocity, \textit{U}\textsubscript{\textrm{\textit{res}}} and drop sizes}}\label{Sec4}
\begin{figure*}[htp!]
   \centering
   \vspace{0pt}
    \includegraphics[width=\columnwidth]{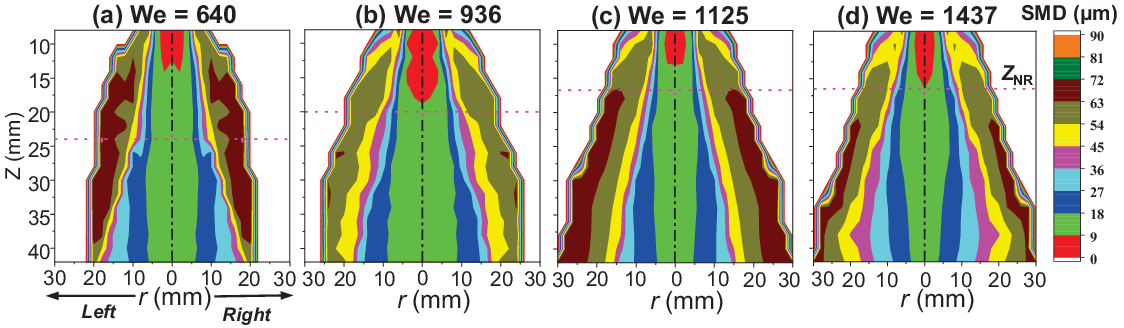}
   \vspace{-15pt}
  \caption{\label{Fig5a}Contour plots of Sauter Mean Diameter, SMD  obtained using PDI for the spray with different \textit{We} from the swirl atomizer. The measurements using PDI are obtained at different spatial locations in right side of the spray axis (shown as vertical black dot dash line) and the data shown in the left side of the spray axis is the same assuming that the spray from the atomizer is axisymmetric. The location $Z_{\textrm{\textit{NR}}}$ is highlighted as pink dashed line.}
\end{figure*}

\begin{figure*}[htp!]
   \centering
   \vspace{0pt}
    \includegraphics[width=\columnwidth]{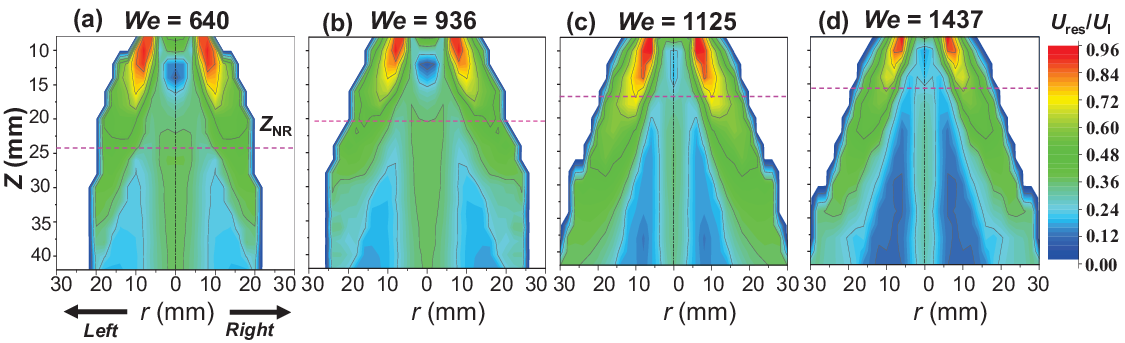}
   \vspace{-15pt}
  \caption{\label{Fig5b}Contour plots of normalized droplet resultant velocity, $U_{\textrm{\textit{res}}/U_l}$ obtained using PDI for the spray with different \textit{We} from the swirl atomizer. The measurements using PDI are obtained at different spatial locations in right side of the spray axis (shown as vertical black dot dash line) and the data shown in the left side of the spray axis is the same assuming that the spray from the atomizer is axisymmetric. The location $Z_{\textrm{\textit{NR}}}$ is highlighted as pink dashed lines.}
\end{figure*}

To understand why the scaled axial velocity shows a marked variation from near and far regions as described in the previous section, herein we present SMD and resultant velocity, $U_{\textrm{\textit{res}}}/U_l$ contour plots. The measurements are obtained at different spatial locations in one side (right) of the spray and reflected on the other side (left) assuming that the spray from the atomizer is axisymmetric. The line demarcating the near and far region of the liquid film breakup is highlighted in Figs. \ref{Fig5a} and \ref{Fig5b} as pink dashed line corresponding to $Z = Z_{NR}$. 
Fig. \ref{Fig5a} \textcolor{black}{(a)}-\textcolor{black}{(d)} shows the contour plots of SMD for the spray with different \textit{We} corresponding to increasing $\Delta P$ from the swirl atomizer. The spatial distribution of SMD is shown in Fig. \ref{Fig5a} as step contours instead of smooth contours to understand the effect of droplet dispersion, secondary breakup, and droplet-droplet coalescence at different spatial locations of \textit{Z} and \textit{r}. It can be seen from Fig. \ref{Fig5a} that the droplets of a single droplet size class of 9 -18 $\mu\textrm{m}$ are seen in the spray axis region, except at $Z$ locations close to the orifice exit. The single droplet size class in the spray axis region shows the absence of droplet breakup (secondary atomization) in this region \textcolor{black}{as justified by our measurements shown in \hypersetup{linkcolor=black}{\ref{SecAppB}}}. 

In the near region of liquid film breakup (well above the black dashed line), the SMD increases initially with increase in \textit{r}, reaches maximum, and then decreases towards the spray periphery, particularly for the spray with high \textit{We} $( > 936 )$. It can also be observed that most spray droplets in the region (above the black dashed line) belong in a single size class, as shown by a single dominant color patch of the contour plots in the region. In the far region of liquid film breakup (below the black dashed line), the SMD increases with \textit{r} and reaches maximum at the spray boundary. Along the spray periphery, SMD increases away from the orifice exit up to a certain $Z$, from there, it either decreases or remains constant, for all \textit{We}. At farther locations from the orifice exit (well below the black dashed line), a significant variation in the size class of the spray droplets is seen, as shown by different color patches in the contour plots. These observed trends in SMD and $U_{\textrm{\textit{res}}}/U_l$ (as described below) could serve as characteristic features to identify the near and far regions of liquid film breakup in the spray.

Fig. \ref{Fig5b}\textcolor{black}{(a)}-\textcolor{black}{(d)} shows the contour plots of normalized droplet mean resultant velocity, $U_{\textrm{\textit{res}}}/U_l$, estimated from the three components of droplet velocities measured from PDI as $U_{\textrm{\textit{res}}}/U_l=\sqrt{U_a^2+U_r^2+U_y^2}/U_l$, (see Fig. \ref{Fig2}) for the spray with different \textit{We} from the swirl atomizer. The spatial variation of normalized $U_{\textrm{\textit{res}}}/U_l$ exhibits different trends above and below the demarcation line. In the near-region of the liquid film breakup, for a given $Z$, the value of $U_{\textrm{\textit{res}}}/U_l$ increases with $r$, reaches maximum, and decreases thereafter moving towards the spray boundary. For the tested spray conditions, the region above the demarcation line encloses the major portion of the recirculatory vortex present in the spray. In the far region of the liquid film breakup (below the dashed line), for a given $Z$, the value of $U_{\textrm{\textit{res}}}/U_l$ decreases with $r$ initially and then increases until it reaches maximum, thereafter, decreases with increase in $r$ until spray boundary, for the spray with $We =$ 1125 and 1437 at all $Z$. But for the spray with $We = $ 640 and 936, similar variation as at high liquid flow rate (or $\Delta P$, see Table \ref{Table1}) is only seen up to a certain $Z$ and then $U_{\textrm{\textit{res}}}/U_l$ decreases with $r$ continuously towards the spray boundary. This spray behavior in the far-region may be attributed to the influence of external factors in the dynamics of the spray.

From Fig. \ref{Fig5b} we can infer that in the far-region of liquid film breakup (beyond $Z_{NR}$), a high-velocity stream is seen along the spray axis for all \textit{We}. The high-velocity stream in the present context is defined as the continuous region of high velocity spray droplets in any direction with a constant $U_{\textrm{\textit{res}}}/U_l$ range, which is approximately 6 - 12 m/s in the current work. With increase in \textit{We}, the width of the high-velocity stream decreases. Additionally, it can be inferred from Fig. \ref{Fig5b} that $U_{\textrm{\textit{res}}}/U_l$ is spatially varies with its maximum value near the orifice exit, $Z = 0$ \citep{Durdina2014}. Similar velocity field is observed in earlier works \citep{Durdina2014, Du2017,Jedelsky2018} of hollow cone swirl sprays. For the spray with $We =$ 1124 and 1437, a high-velocity stream is seen along the spray periphery apart from the one along the spray axis. An interesting observation to be noted is that the high-velocity stream approximately originates at the demarcation line ($Z_{NR}$) at all \textit{We}.

It can be seen from Table \ref{Table2} and the contour plots (see Figs. \ref{Fig5a} and \ref{Fig5b}) that value of $Z_{NR}$ varies with \textit{We} and the effect of liquid film breakup on the drop size characteristics is seen for smaller distances from the orifice exit with \textit{We}. This implies that the value of $Z_{NR}$ depends on the variation of $L_b$. \textcolor{black}{Thus, it can be concluded, for the range of selected operating parameters corresponding to our injector, the near region of liquid film breakup where film breakup influences drop size characteristics is approximately located 2 to 2.5 times $L_b$ from the orifice exit.} Since the behavior of spray droplet characteristics is different in the near region and far region of the liquid film breakup, a comparison of spray droplet characteristics for these two regions provides more precise insights on the final droplet size and drop size distribution for the swirl atomizer. Against this backdrop we select two axial locations are selected to analyze these characteristics: $Z = Z_b$ for the near-region of the liquid film breakup and $Z =$ 30 mm for the far region of the liquid film breakup for further analysis.

\subsection{\textbf{Number and volume flux distributions in near and far regions of the spray}}\label{Sec5}
\begin{figure*}[htp!]
   \centering
   \vspace{0pt}
    \includegraphics[scale =0.8]{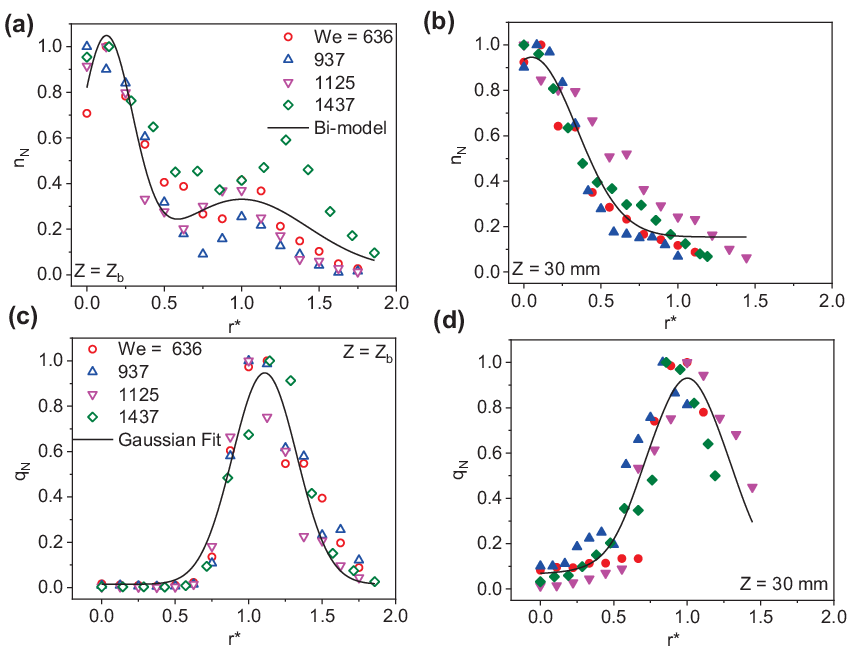}
   \vspace{-15pt}
  \caption{\label{Fig6}Variation of normalized number flux and normalized volume flux with $r^{*}$ for the spray from the swirl atomizer with different \textit{We} at $Z = Z_b$ and $Z = 30$ mm. The details of the unimodal and bimodal Gaussian fits shown as solid black lines in these plots are provided in \ref{SecAppC}.}
\end{figure*}

Our choice of $r_{mvf}$ was based on the considerations of volume flux and therefore we describe these in detail for near and far regions of the spray in this section. Figure \ref{Fig6} \textcolor{black}{(a)} - \textcolor{black}{(d)} shows the variation of normalized number flux, $n_N$ and volume flux, $q_N$ with $r^{*}$ for the spray with different \textit{We} in the near ($Z = Z_b$) and far-region of the spray ($Z = 30$ mm). The droplet number flux deduced from the PDI measurements provides the details about the droplet concentration at given measurement location (\textit{Z}, \textit{r}) in the spray. 
The number flux at $(Z, r)$, $n$, is obtained by dividing the total droplet count with the product of the data acquisition time, $t_{aq}$ and the Ch1 (channel 1) validation. It is expressed as,
\begin{equation}
n = \dfrac{\textrm{Count}}{t_{aq} \times \textrm{Ch1 validation}}
\end{equation}
Here, the parameter “Ch1 validation” is a measure of spray droplets validated by PDI as being completely within the probe volume and giving a reliable measurement. The normalized number flux, $n_N$ at a given measurement location in the present study is obtained by dividing the number flux at a given ($Z$, $r$) with the maximum number flux for that axial location, $Z$, and is expressed as,
\begin{equation}
n_N = \dfrac{n}{n_{max}}
\end{equation}
The volume flux can be directly obtained from PDI, and the normalized volume flux, $q_N$ is obtained in the similar way as the normalized number flux. It can be observed from Fig. \ref{Fig6}\textcolor{black}{(a)} and \ref{Fig6}\textcolor{black}{(c)} that the maximum $n_N$ is seen at the spray axis and the maximum volume flux is seen at the film breakup zone. This shows that the region of spray axis consists of large number of small droplets, and the breakup zone is filled with the larger droplets which are produced from the breakup of liquid film \citep{Hinterbichler2020} which results in two distinct peaks in Fig. \ref{Fig6}\textcolor{black}{(a)}.

In the far-region of the spray, the second smaller peak at the breakup zone (see Fig. \ref{Fig6}\textcolor{black}{(b)}) is completely absent. In other words, the memory of liquid film breakup is not carried in the far region of the liquid film breakup in the spray \citep{Rajamanickam2019}. Another characteristic feature of the far region of the spray is seen in Fig.\ref{Fig6}\textcolor{black}{(d)}. The maximum volume flux of the spray at a given $Z$ in the far-region is seen closer to the spray boundary (see Fig. \ref{Fig3}), possibly due to the centrifuge effect of the spray droplets \citep{Saha2012}.

\subsection{\textbf{Drop size and velocity correlation in near, $Z = Z_b$ and far regions, $Z =$ 30 mm}}\label{Sec6}
\begin{figure*}[htp!]
   \centering
   \vspace{0pt}
    \includegraphics[width=\textwidth]{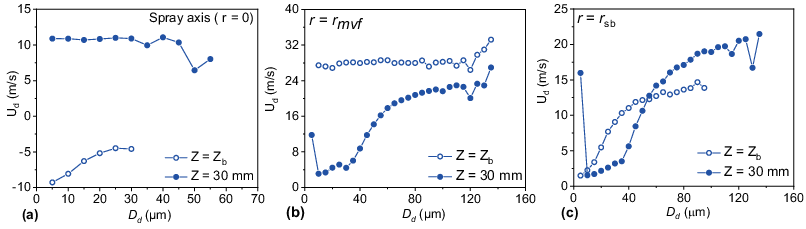}
   \vspace{-15pt}
  \caption{\label{Fig7} Droplet diameter ($D_d$) - axial velocity ($U_a$) correlation at $Z = Z_b$ and $Z = 30$ mm for the spray from the swirl atomizer with $We = 1437$ at different radial locations. (a) $r = 0$, (b) $r = r_{mvf}$, and (c) $r = r_{sb}$.}
\end{figure*}

The number and volume flux distributions described above depend strongly on two factors - the drop diameter and its velocity. To understand the relationship between droplet diameter ($D_d$) and droplet axial velocity ($U_a$), the $D_d - U_a$ correlations at different radial locations are presented in the near and far regions of liquid film breakup for the spray with $We =$ 1437 in Fig. \ref{Fig7}. 
It can be observed from Fig. \ref{Fig7}(a) that at $r = 0$, $U_a$ has a negative value at $Z = Z_b$ but increases  towards positive values with increase in $D_d$, but at $Z =$ 30 mm, $U_a$ maintains constant value with increase in $D_d$ for majority of the size classes and then decreases. The negative value of $U_a$ at $Z = Z_b$ is due to the presence of recirculation vortex in the swirl spray. At $Z =$ 30 mm, the constant value of $U_a$ shows that, irrespective of the size, the droplets move with the same velocity near the spray axis and the same is observed for other \textit{We} too (not shown here). At $r = r_{mvf}$, $U_a$ is independent of $D_d$ at $Z = Z_b$ \citep{Vankeswaram2022}, but $U_a$ continuously increases with $D_d$ at $Z =$ 30 mm. The same is observed at all \textit{We}(not shown here) for brevity.  At $r = r_{sb}$, for both $Z$ tested in the present case, $U_a$ increases with the increase in $D_d$ which may be attributed to higher inertia/weight of the drops by virtue of them being larger in size.

The results given in Fig. \ref{Fig7} are the local variations in $D_d - U_a$ correlation in the spray. To understand the global variation in the $D_d - U_a$ correlation, the $\left(D_d - U_a\right)$ global correlation plots are obtained for a given $Z$ using all the local measurements (radial locations), which are estimated using the using the following expression,
\begin{equation}
{\left(D_d - U_a\right)}_{global} = \sum_{i=1}^{I} \left\{\sum_{j=0}^{J} \left(\sum_{k=1}^{k_{ij}} U_{a,jk}\right)\dfrac{1}{\sum_{j=0}^{J} X}\right\}
\end{equation}
where $i$ is the droplet size class, $j$ is the radial measurement location, and \textit{X} is the length vector where  $k_{ij} \in X$; $k_{ij}$ corresponds to length of the $i^{th}$ class for $j^{th}$ location.
\begin{figure*}[htp!]
   \centering
   \vspace{0pt}
    \includegraphics[width=\textwidth]{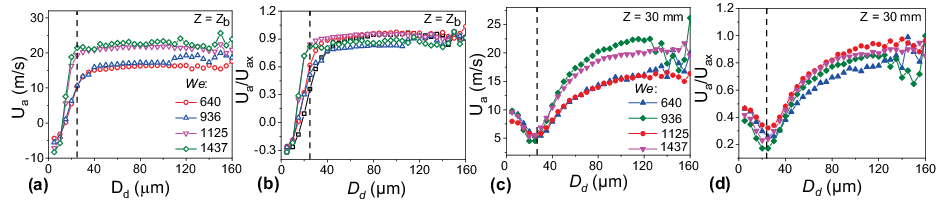}
   \vspace{-15pt}
  \caption{\label{Fig8}Variation of global $D_d - U_a$ correlation for different \textit{We} at $Z = Z_b$ for, (a) Droplet axial velocity ($U_{a}$) (b) Droplet axial velocity ($U_{a}$) scaled by axial sheet velocity ($U_{ax}$) and at $Z =$ 30 mm for, (c) Droplet axial velocity ($U_{a}$) (d) Droplet axial velocity ($U_{a}$) scaled by axial sheet velocity ($U_{ax}$).}
\end{figure*}

It is expected that the droplet velocity increase with increase in drop size in sprays, but at $Z =$ 30 mm, it is seen in the other way for $D_d \le$ 25\;$\mu$m, which is identified as tea-spoon effect by Hinterbichler et al. \citep{Hinterbichler2020} for $D_d <$15 $\mu$m. This can be further explained by using the variations presented in Fig. \ref{Fig7} \textcolor{black}{(a)}-\textcolor{black}{(c)} and Fig. \ref{Fig8}. In the far region of liquid film breakup, a high-speed stream is observed along the spray centerline, where the SMD and volume flux are minimum, and the number flux is maximum. This shows that the small ($D_d < 5\;\mu$m) and a portion of medium ($5\;\mu\textrm{m} \le  D_d \le 25\;\mu\textrm{m}$) droplets (division of size classes is presented further) carried towards the spray axis due to the spray dispersion and air-entrainment process, while the large droplets are carried towards the spray periphery due to their inertia. These droplets due to being exposed to the airflow longer are able to exchange momentum with the gas phase extensively \citep{Park2006}.

Figure \ref{Fig8}\textcolor{black}{(a)}-\textcolor{black}{(d)} shows $\left(D_d - U_a\right)$ correlation (global data) of the spray droplets at values of $Z$ in the near ($Z = Z_b$) and far-region ($Z =$ 30 mm) of the liquid film breakup for the spray from the swirl atomizer with different \textit{We}. At $Z = Z_b$, in Fig. \ref{Fig8} \textcolor{black}{(a)} and \textcolor{black}{(b)} for the spray droplets with $D_d >$ 25 $\mu m$, $U_a$ is independent of $D_d$ for all \textit{We} (highlighted by black dashed line), and for $D_d \le$ 25 $\mu\textrm{m}$, the value of $U_a$ increases for all \textit{We}. At $Z = 30$ mm, for $D_d \le$ 25\;$\mu\textrm{m}$, the value of $U_a$ decreases for all \textit{We} and for $ D_d >$ 25\;$\mu\textrm{m}$, $U_a$ continuously increases with $D_d$ as seen from Fig. \ref{Fig8}\textcolor{black}{(c)} and \textcolor{black}{(d)}. Also one may notice that the $U_a - D_d$ correlation (global data) exhibits two peaks for the spray at $Z = 30$ mm as opposed to a single peak at $Z = Z_b$ for all \textit{We}. 

Furthermore, from Figs. \ref{Fig8}\textcolor{black}{(c)} and \textcolor{black}{(d)} the axial velocity of the spray droplets scales well by the liquid sheet axial velocity ($U_a$) as the measurements of droplet velocity normalized with $U_a$ collapse into a single trend.  These observations provide us a good idea on the $\left(D_d - U_a\right)$ correlations as we mover from near and far regions and the effect of film breakup diminishes and other effects such as secondary atomization and drop ballistics become significant. 

\subsection{\textbf{Global (mean) drop size characteristics}}\label{Sec7}
\begin{figure*}[htp!]
   \centering
   \vspace{0pt}
    \includegraphics[scale=1.2]{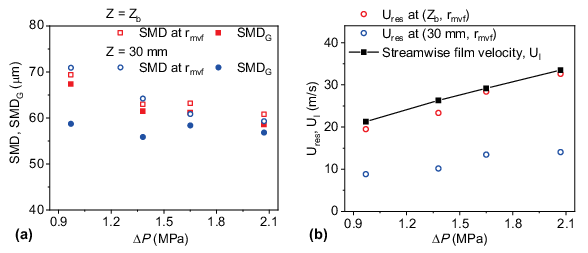}
   \vspace{-15pt}
  \caption{\label{Fig9} Comparison at \textit{r\textsubscript{mvf}} of, (a) SMD at $Z = Z_b$, and $Z = 30$ mm with $\textrm{SMD}_{\textrm{G}}$ obtained using Eq. \ref{globalSMD} for different \textit{We}, and (b) droplet mean resultant velocity at $Z = Z_b$, and $Z = 30$ mm with the liquid sheet streamwise velocity.}
\end{figure*}
The spray at the radial location corresponding to the droplet maximum volume flux represents the global spray for the corresponding axial measurement location \citep{Vankeswaram2022}. Evidently, the SMD is a useful measure in understanding the expected average drop sizes at various axial locations. Larger drop sizes are indicative of higher volume flux but a lower number flux and therefore can help us understand spray flux data better. In this spirit, an attempt is made to compare the SMD at $r_{mvf}$ with global Sauter mean diameter, SMD\textsubscript{G} obtained using Eq. \ref{globalSMD} given by Tratnig and Brenn \citep{Tratnig2010} at $Z = Z_b$ and $Z =$ 30 mm for different \textit{We}. The global SMD, SMD\textsubscript{G} at different \textit{Z} is mathematically represented by,
\begin{equation}\label{globalSMD}
\textrm{SMD}_{\textrm{G}} = \dfrac{\sum\limits_{j=1}^{J}\sum\limits_{i=1}^{I}D_{d,i}^3(r_j)\;n(r_j,D_{d,i})r_j}{\sum\limits_{j=1}^{J}\sum\limits_{i=1}^{I} D_{d,i}^2(r_j)\;n(r_j,D_{d,i})r_j}
\end{equation}
Here subscripts $i$ and $j$ refer, respectively, to the number of size classes in the drop size distribution and the number of radial positions recorded by the PDI, $r_j$ is the radial distance of the measurement position $j$, $D_{d,i}(r_j)$ is the mean droplet size of the size class $i$ at the measurement position $r_j$, and $n(r_j,D_{d,i})$ denotes the number flux of the droplets with sizes in the size class $i$ at the measurement position $r_j$. The value of $n(r_j,D_{d,i})$ is obtained by dividing the droplet rate by the respective Ch1 validation. An excellent match is observed from Fig. \ref{Fig9}\textcolor{black}{(a)} at $Z = Z_b$ between the estimated value of $\textrm{SMD}_{\textrm{G}}$ and SMD at $r = r_{mvf}$ \citep{Vankeswaram2022}. Whereas at Z = 30 mm, a close match is seen for high \textit{We}, but for low \textit{We}, the SMD at $r = r_{mvf}$ is higher than the estimated value of $\textrm{SMD}_{\textrm{G}}$. 

These observed differences  seen for the spray with low \textit{We} in the far-region could be mainly due to the decrease in radial dispersion of the spray at $Z$ far from the liquid film breakup as seen in Fig. \ref{Fig2}\textcolor{black}{(b)}; with increase in $Z$, the $r_{mvf}$ is moving towards the spray axis for low \textit{We}, but it is followed the spray periphery for high \textit{We}. Figure \ref{Fig9}\textcolor{black}{(b)} shows the comparison of droplet mean resultant velocity at $r = r_{mvf}$ with the liquid sheet streamwise velocity, $U_l$ for different \textit{We} at two different axial locations. It can be observed from Fig. \ref{Fig9}\textcolor{black}{(b)} that an excellent match is seen at $Z = Z_b$ \citep{Vankeswaram2022}, but a poor match is observed at $Z =$ 30 mm. With increase in axial distance from liquid film breakup the interaction of spray droplets with the surrounding ambient air significantly decelerates the drops \cite{Saha2012,Durdina2014}, this decrease in droplet momentum is the reason for the poor match at $Z =$ 30 mm. The analysis presented in Fig. \ref{Fig9} identifies that the spray at $r = r_{mvf}$ can only be the representative sample of the global spray in the near region of the liquid film breakup.

\subsection{\textbf{Drop size distribution in near and far regions}}\label{Sec8}
In order to explore the underlying statistics leading to the observed mean drops sizes shown in the previous section we delve deeper into the global drop size distributions (at a particular \textit{Z} considering data from all \textit{r}). It is shown in previous studies that global spray characteristics are a good direct measure of the performance of an atomizer \citep{Tratnig2010,Dhivyaraja2019}. Traitng and Brenn\citep{Tratnig2010} proposed a global drop size probability distribution function (PDF) for their experimental data and given by,
\begin{equation}\label{eqnPDF}
\textrm{PDF} = \dfrac{1}{\Delta D_d}\dfrac{\sum\limits_{j=1}^{J}n(r_j,D_{d,i})r_j}{\sum\limits_{j=1}^{J}\sum\limits_{i=1}^{I}n(r_j,D_{d,i})r_j}
\end{equation} 
Here, suffix $i$ refers to the number of size classes in the drop size distribution, suffix $j$ refers to the number of radial positions recorded by the PDI, $r_j$ is the radial distance of the measurement position $j$, $D_{d,i}(r_j)$ is the mean drop size of the size class $i$ at measurement position $r_j$, and $n(r_j,D_{d,i})$ denotes the number flux of the drops with sizes in size class $i$ at the measurement position $r_j$. The value of $n(r_j,D_{d,i})$ is obtained by dividing the drop rate by the Ch1 validation rate recorded in the PDI measurements. 
\begin{figure*}[htp!]
   \centering
   \vspace{0pt}
    \includegraphics[scale=0.7]{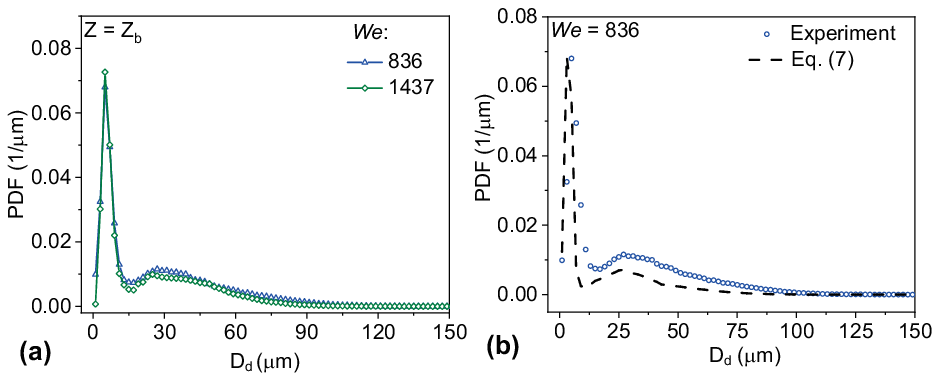}
   \vspace{-10pt}
  \caption{\label{Fig10}(a) The variation of experimental global drop size PDF at $Z = Z_b$ for the spray with different \textit{We} from the swirl atomizer, and (b) the comparison of the experimental global drop size PDF with the prediction obtained using Eq. \ref{combtwo} for the spray with $We = 836$ from the swirl atomizer.}
\end{figure*}
The value $\Delta D_d$ is 2 $\mu\textrm{m}$ in the present case. Figure \ref{Fig10}\textcolor{black}{(a)} shows the PDF of global drop size distribution, obtained using Eq. \ref{eqnPDF} with the value of $\Delta D$ taken as 2 $\mu\textrm{m}$, at $Z = Z_b$ for different \textit{We}. It can be observed from Fig. \ref{Fig10}\textcolor{black}{(a)} that the global drop size distribution at $Z = Z_b$ exhibits a bimodal distribution with peaks at two different values of $D_d$. This variation can be attributed to the increase in the production of number of satellite droplets with \textit{We} \citep{Park2006}.

For bimodal distributions, such as the one seen here, a combination of two normal distributions may be used \citep{Park2010}. 
\begin{equation}\label{combtwo}
f (D_d)=\dfrac{P}{S_{N,1}\sqrt{2\pi}}\;e^{-\frac{1}{2}\left(\frac{D_d-\overline{D}_1}{S_{N,1}}\right)^2} + \dfrac{1-P}{S_{N,2}\sqrt{2\pi}}\;e^{-\frac{1}{2}\left(\frac{D_d-\overline{D}_2}{S_{N,2}}\right)^2}                                                                     
\end{equation}
where, $S_{N,1}$ and $\overline{D}_1$ are the deviation and mean values for first normal distribution and $S_{N,2}$ and $\overline{D}_2$ are for the second normal distribution. The variable $P$ is the value for normalization. If $N$ is the total number of droplets presented in the distribution and $N_1$ and $N_2$ is the droplets share by the two normal distributions, then $P$ is obtained by, $N_1/N$. \textcolor{black}{For the \textit{We} in the range $630 - 1147$ we compute the following values for the fit parameters, $P = 0.65 – 0.43$, $S_{N,1} = 3.25 – 2.12$, $S_{N,2} = 20.9 – 16.5$, $\overline{D}_1 = 11.03 – 7.1 $, and $\overline{D}_2 = 34.5 – 29.8$.}

The bimodal distribution can be represented by using a combination of two normal distributions \citep{Park2006} as given in Eq. \ref{combtwo}. To compare the experimental drop size distribution with the function given by Eq. \ref{combtwo}, the drop size spectra is divided into two groups; one is from 0$-$\textit{R} and the other is $>$ \textit{R} (where, \textit{R} is the range of the group). This range is obtained through trial and error to obtain best possible fit. The values of $S_{N,1}$, $D_1$, $S_{N,2}$ and $D_2$ are obtained by using the respective drop size distribution. It can be observed from Fig. \ref{Fig10}\textcolor{black}{(b)} that a satisfactory match is seen between the experimental global PDF with the one estimated using Eq.\ref{combtwo} though the theory under predicts the second peak.

\begin{figure*}[htp!]
   \centering
   \vspace{0pt}
    \includegraphics[scale=0.7]{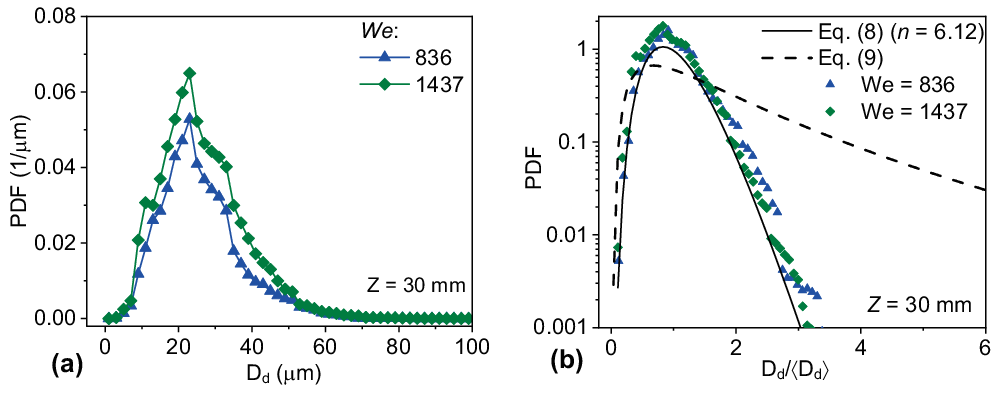}
   \vspace{-10pt}
  \caption{\label{Fig11}(a) The variation of global drop size PDF at $Z = 30$ mm for the spray with different \textit{We} from the swirl atomizer, and (b) comparison of the experimental global drop size PDF with the spray drop size distribution models of gamma distribution and log-normal distribution.}
\end{figure*}

Figure \ref{Fig11}\textcolor{black}{(a)} shows the PDF of global drop size distribution, obtained using Eq. \ref{eqnPDF} with the value of $\Delta D_d$ taken as 2 $\mu\textrm{m}$, in the far region of the spray, at $Z = 30$ mm for different \textit{We}. It can be observed from Fig. \ref{Fig11}\textcolor{black}{(a)} that the global drop size PDF at $Z =$ 30 exhibits unimodal distribution, for all \textit{We}. For such sprays dominated by ligament mediated drop formation, the probability density function (PDF), $P_b$ can be described by the gamma function given as \citep{Villermaux2004, Kooij2018}, 
\begin{equation}\label{gammadis}
P_b \left(\alpha, x = \dfrac{D_d}{\langle D_d \rangle}\right)= \dfrac{\alpha^\alpha}{\Gamma(\alpha)}x^{\alpha-1} e^{-\alpha x}
\end{equation}
where $\langle D_d \rangle$ is the average droplet diameter, and $\alpha$ is the gamma distribution ($\Gamma$) parameter. For a given spray, $\alpha$ is determined experimentally from the characteristics of ligaments formed in the spray \citep{Vankeswaram2022}. 
The value of $\alpha$ is found to be 4 to 5 for very corrugated ligaments, and $\alpha = \infty$ for the smoothest ligaments \citep{Villermaux2004, Kooij2018}. A value of $\alpha = 6.12$ is used in the present case \citep{Vankeswaram2022}. Another choice to fit data for a uni-modal distribution is the Log-normal function which provides reasonable fit to many particle/drop size distributions that occur in nature especially close to their maximum \citep{Lefebvre2017, Kooij2018}. It is mathematically expressed as,
\begin{equation}\label{lognorm}
\varphi =  \dfrac{1}{\sqrt{2\pi}Ds_g} e^{-\dfrac{1}{2s_g^2} \textrm{\normalsize log } \left(\dfrac{D_d}{D_{n_g}}\right)^2}
\end{equation}
where $D_{n_g}$ is the number mean drop size and $s_g$ is the standard deviation. 

Figure \ref{Fig11}\textcolor{black}{(b)} shows the comparison of experimental global drop size PDF obtained using Eq. \ref{eqnPDF} with gamma distribution (Eq. \ref{gammadis}) and log-normal distribution (Eq. \ref{lognorm}). It can be seen from Fig.\ref{Fig11}\textcolor{black}{(b)} that gamma distribution predicts the global PDF well unlike log-normal distribution. Lan et al. \citep{Lan2014} compared experimental data obtained using Eq. \ref{eqnPDF} with Rosin-Rammler, Nukiyama-Tanasawa and found a poor match with them. So far, no literature is available for the comparison of gamma distribution using Eq. \ref{gammadis}. Although, Kooij et al., \citep{Kooij2018} found a good match with both gamma distribution and log-normal distribution their global drop size distribution data was directly measured using Spraytec\textsuperscript{\textregistered} unlike our case.

It is shown earlier that the mathematical model of gamma distribution (Eq. \ref{gammadis}) with a value of $\alpha = 6.12$ predicted the drop size distribution at a particular measurement location (local PDF) very well \citep{Vankeswaram2022}. Here, we observe the same for our global drop size distribution experimental data obtained using Eq. \ref{combtwo}, wherein the gamma distribution (Eq. \ref{gammadis}) fits it well as shown in Fig. \ref{Fig11} \textcolor{black}{(b)}. Comparatively from Fig. \ref{Fig11} \textcolor{black}{(b)} it is seen the log-normal distribution obtained using Eq. \ref{lognorm} fails to predict the experimental drop size distribution with any reasonable accuracy.

\subsection{\textbf{Axial velocity characteristics in near and far region based on drop size classification}}\label{Sec9}
\begin{figure*}[htp!]
   \centering
   \vspace{0pt}
    \includegraphics[scale=1.0]{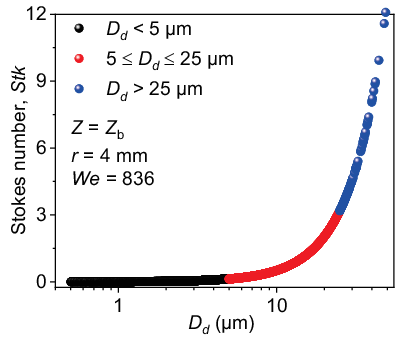}
   \vspace{-15pt}
  \caption{\label{Fig12}Variation of \textit{Stk} for different size classes of the droplets for the spray at ($Z$, $r$) = ($Z_b$, 4 mm) from the swirl atomizer.}
\end{figure*}
The recirculation zone created due to the dynamic interaction of the liquid spray with the surrounding air flow in a pressure swirl atomizer plays a crucial role in the air entrainment process which enhances ignition and flame stabilization \citep{Rosa1992, Steinberg2013, Elbadawy2015}. \textcolor{black}{The presence of small droplets near the spray axis and large ones near the spray periphery are indicative of the effect of the recirculation zone  \hypersetup{citecolor=black}{\citep{Belhadef2012}}. Our measurements too show a similar trend and we take that as a confirmation of the presence of zone in our spray too.} Since a spray is an ensemble of drops of different sizes which behave differently due to varying inertia and their ability to follow the surrounding air flow, their spray characteristics in groups of specific drop diameter can provide crucial insights into the underlying physical process of entertainment. 

A convenient to sort the drop in different bin sizes is by use of the dimensionaless Stokes number, \textit{Stk} which is an indicator of the level of responsiveness of individual droplets to the turbulent fluctuations of the surrounding gas phase is defined from the ratio of droplet relaxation time, $\tau_d$ to the turbulent time scale \citep{Sahu2016},  $\tau_t$ as $\textrm{\textit{Stk}} = \tau_d/\tau_t$ where $\tau_d = \frac{\rho_d D_d^2}{18\mu_g}$,  $\tau_t=l_c/U_{\textrm{RMS}}$, $\rho_d$ is the droplet density (liquid density), $D_d$ is the droplet diameter, $\mu_g$ is the gas density, $U_{\textrm{RMS}}$ is the RMS (root mean squared) variation of droplet axial velocity, and $l_c$ is the turbulent length scale, which is defined as $1/5$\textsuperscript{th} of the spray width \citep{Kooij2018}. 

Simply speaking, a droplet with a low \textit{Stk} follows the air flow perfectly and on the other hand droplets with a large \textit{Stk} is dominated by inertia and follows its own prescribed path. In the current work, the characteristics of surrounding gaseous phase are obtained from the spray droplet characteristics deduced from the PDI. It is assumed that the droplets whose \textit{Stk} much less than 1 faithfully follow the surrounding air flow \citep{Jedelsky2018, Sahu2016} and therefore the droplets whose \textit{Stk} $<< 1$ can be used to represent the gas phase. Based on the results obtained in previous studies \citep{Jedelsky2018, Sahu2016} and the values of \textit{Stk} in the present case, the spray droplets are divided into 3 size classes: small, S ($D_d < 5\;\mu\textrm{m}$; \textit{Stk} $<<$ 1), medium, M ($5\;\mu\textrm{m} \le D_d \le 25\;\mu\textrm{m}$; \textit{Stk} $\sim O(1)$), and large, L ($D_d > 25\;\mu\textrm{m}$; \textit{Stk} $>>$ 1). 
\begin{figure*}[t!]
   \centering
   \vspace{0pt}
    \includegraphics[width=\textwidth]{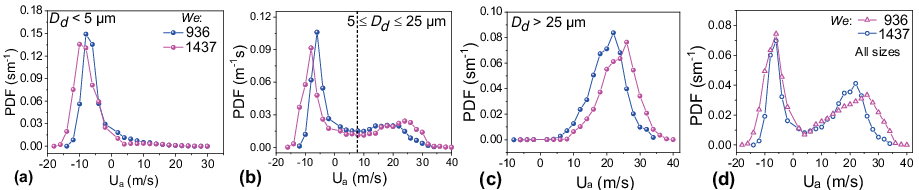}
   \vspace{-15pt}
  \caption{\label{Fig13}Variation of global droplet axial velocity PDF for different size classes at $Z = Z_b$ for the spray with different \textit{We} from the swirl atomizer. (a) $D_d < 5\;\mu\textrm{m}$ (b) $5\;\mu\textrm{m} \le D_d \le 25\;\mu\textrm{m}$, and (c) $D_d > 25\;\mu\textrm{m}$. (d) Variation of global droplet axial velocity PDF for all three size classes combined at $Z = Z_b$ for the spray with different \textit{We} from the swirl atomizer.}
\end{figure*}

Figure \ref{Fig12} shows the variation of \textit{Stk} with $D_d$ obtained at $r =$ 4 mm and $Z = Z_b$ for the spray with $We =$ 936 from the swirl atomizer. The spray droplets of size less than 5 $\mu m$ (S) are having \textit{Stk} much less than 1 (black filled circles). For $5\;\mu\textrm{m} \le D_d  \le 25\;\mu\textrm{m}$ (M), the value of \textit{Stk} is O(1) (red filled circles). For $D_d > 25\;\mu\textrm{m}$ (L), the value of \textit{Stk} is much greater than 1 (blue filled circles). From the value of \textit{Stk} given in Fig. \ref{Fig12} for the spray droplets, the mean value, $\langle \textrm{\textit{Stk}} \rangle$ is obtained for different size classes. Considering the radial variation of $\langle \textrm{\textit{Stk}} \rangle$ for the different size classes at $Z = Z_b$ and $Z =$ 30 mm for the spray with different \textit{We} (not shown here), the small droplets are having $\langle \textrm{\textit{Stk}} \rangle$ much less than 1 at all $r$, and for the medium droplets, the value of $\langle \textrm{\textit{Stk}} \rangle$ is less than 1 at \textit{r} close to the spray axis and is very close to 1 at other \textit{r}. 
\begin{figure*}[b!]
   \centering
   \vspace{0pt}
    \includegraphics[width=\textwidth]{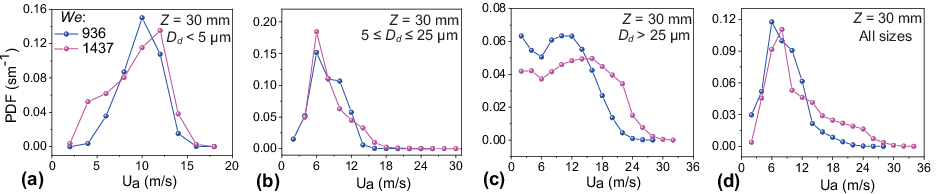}
   \vspace{-15pt}
  \caption{\label{Fig14}Variation of global droplet axial velocity PDF for different size classes at $Z = 30$ mm for the spray with different \textit{We} from the swirl atomizer. (a) $D_d < 5\;\mu\textrm{m}$ (b) $5\;\mu\textrm{m}  \le D_d \le 25\;\mu\textrm{m}$, and (c) $D_d > 25\;\mu\textrm{m}$. (d) Variation of global droplet axial velocity PDF for all three size classes combined in the far-region at $Z = 30$ mm for the spray with different \textit{We} from the swirl atomizer.}
\end{figure*}

This shows that the medium size class too can be influenced by the surrounding airflow \citep{Jedelsky2018}. For the large size class, the value of $\langle \textrm{\textit{Stk}} \rangle$ is nearly 1 at the r close to the spray axis and is much greater than 1 at other \textit{r}. This indicates that this size class is least effected by the surrounding airflow. This reveals the significance of setting the threshold droplet size to ensure proper representation of the gas phase flow field by the experimentally measured droplet velocities \citep{Hinterbichler2020}. Thus, all the small size class droplets and a portion of medium size class droplets are used to represent the gas phase in the present case. To identify the droplet size threshold for the medium size class, a similar approach as given in Hinterbichler et al. \citep{Hinterbichler2020} is followed. Therefore, much like Eq. \ref{globalSMD} the PDF of global velocity distribution of spray droplets can be obtained for the droplet axial velocity ($U_a$) using the following expression,
\begin{equation}
\textrm{PDF} = \dfrac{1}{\Delta U_a}\dfrac{\sum\limits_{i=1}^{J}n(r_j,U_{a,i})r_j}{\sum\limits_{i=1}^{J}\sum\limits_{i=1}^{I}n(r_j,U_{a,i})r_j}
\end{equation} 
In the above suffix \textit{i} refers to the number of velocity classes in the velocity distribution, suffix \textit{j} refers to the number of radial positions recorded by the PDI, $r_j$ is the radial distance of the measurement position \textit{j}, and $n(r_j,U_{a,i})$ denotes the number flux of the drops with velocities in velocity class \textit{i} at the measurement position $r_j$. The value of $n(r_j,U_{a,i})$ is obtained by dividing the drop rate by the Ch1 validation rate recorded in the PDI measurements. The value $\Delta U_a$ is taken as 2 m/s in the present case.

Figure \ref{Fig13} shows the PDF of the global axial velocity distribution of spray droplets of different size classes at $Z = Z_b$ for the spray with different \textit{We} from the swirl atomizer. For the small and medium droplet size classes (see Fig. \ref{Fig13}\textcolor{black}{(a)} and \textcolor{black}{(b)}), the peak of the distribution is seen on the negative velocity side of the distribution, and for the large size class (see Fig. \ref{Fig13}\textcolor{black}{(c)}), it is seen on the positive side. The different peak levels of different droplet size classes in the global PDF data result in the bi-modal distribution in the global axial velocity distribution of the spray droplets (see Fig. \ref{Fig13}\textcolor{black}{(d)}). The first peak is seen almost at the same droplet velocity for both \textit{We} and it is on the negative side of the droplet velocity, but the other peak is moving towards the higher values of the droplet velocity with \textit{We}. 

A small peak is seen on the positive side of $U$ for the medium size class, where the global PDF exhibits bi-modal distribution for this size class. The local minimum between the two peaks of the global PDF for the medium size class (which is different for different \textit{We}), marked by the dashed vertical line, is defined as the threshold velocity \citep{Hinterbichler2020}. Thus, the droplets whose velocity less than the threshold velocity are considered to represent the gas phase for the medium size class. Henceforth, all the droplets in the smallest size class, and the droplets whose velocity less than threshold velocity are used to represent the gas phase flow in the present case. Also, it is to be noted that unlike at $Z = Z_b$, bi-modal global PDF is not seen for any class size as shown in Fig.\ref{Fig14} for $Z =$ 30 mm. 

It can be observed from Fig. \ref{Fig14}\textcolor{black}{(a)}-\textcolor{black}{(d)} that for all size classes the peak of the distribution is seen on the positive side of the droplet axial velocity unlike at $Z = Z_b$ as seen in Fig. \ref{Fig13}\textcolor{black}{(a)}-\textcolor{black}{(d)}. For the small and medium size classes (see Fig. \ref{Fig14}\textcolor{black}{(a)} and \textcolor{black}{(b)}) , not much difference is seen in the peak of the distribution, but for the large size class ((see Fig. \ref{Fig14}\textcolor{black}{(d)} and \textcolor{black}{(b)})) the peak of the distribution shift towards the higher droplet axial velocity with \textit{We}. The same peak levels of different droplet size classes in the global PDF data result in the unimodal distribution in the global axial velocity distribution of the spray droplets. Figure \ref{Fig14}\textcolor{black}{(d)} shows the global PDF of droplet axial velocity for the combined size class at $Z =$ 30 mm for the spray with different \textit{We} from the swirl atomizer. As indicated earlier, the global axial velocity PDF at $Z =$ 30 mm exhibits unimodal distribution. Not much difference is observed in global PDF with \textit{We} at $Z =$ 30 mm.
\begin{figure*}[htp!]
   \centering
   \vspace{0pt}
    \includegraphics[width=\textwidth]{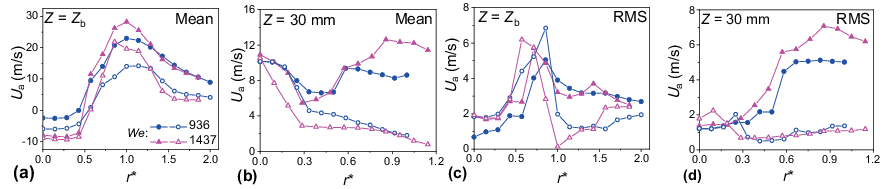}
   \vspace{-15pt}
  \caption{\label{Fig15}Variation of the mean axial velocity for gaseous (unfilled symbols) and liquid phases (filled symbols) along the radial distance at (a) $Z = Z_b$ and (b) $Z =$ 30 mm for the spray with different \textit{We} from the swirl atomizer. Variation of the RMS axial velocity, $U$ for gaseous (unfilled symbols) and liquid phases (filled symbols) along the radial distance at (c) $Z = Z_b$ and (d) $Z = 30$ mm for the spray with different \textit{We} from the swirl atomizer.}
\end{figure*}

Figure \ref{Fig15}\textcolor{black}{(a)}-\textcolor{black}{(d)} shows the mean and RMS (root mean squared) variation of droplet axial velocity for the liquid and gas phase at $Z = Z_b$ (see Fig. \ref{Fig15}\textcolor{black}{(a)}) and $Z =$ 30 mm (see Fig. \ref{Fig15}\textcolor{black}{(b)}). At $Z = Z_b$, no difference in the mean variation of \textit{U} is seen for the liquid and gaseous phases at all $r^{*}$, though the liquid phase is having higher mean values than the gaseous phase. However, at $Z = 30$ mm, the liquid and gaseous phases exhibit different behavior at all $r^{*}$, except at $r^{*}$ close to the spray axis. For the liquid and gaseous phase, a negative droplet velocity is observed at $r^{*}$ close to the spray axis, which is due to the presence of recirculatory vortex, and these are for the droplets whose $\langle \textrm{\textit{Stk}}\rangle$ values are less than 1. A peak in the droplet axial mean velocity at the spray axis (central high-speed stream) is observed for both liquid and gaseous phases at $Z = 30$ mm \citep{Durdina2014, Jedelsky2018}. 

A similar RMS variation is seen for the liquid and gaseous phases with different \textit{We} at $Z = Z_b$ (see Fig. \ref{Fig15}\textcolor{black}{(c)}) and $Z = 30$ mm (see Fig. \ref{Fig15}\textcolor{black}{(d)}). The peak value of droplet axial RMS velocity is seen at a radial location where the droplet mean axial velocity value changes sign from negative to positive at $Z = Z_b$. Also, these values are higher for the gaseous phase than the liquid phase at $Z = Z_b$.

\subsection{\textbf{Characterizing turbulence in the spray}}\label{Sec10}
\textcolor{black}{For the sprays the exchange between the surrounding airflow (continuous phase) to the liquid (dispersed phase) is determined by the turbulence levels at a particular location. To ascertain that fluctuations measured by PDI velocity data are not random but indeed indicative of turbulent we determine the turbulence intensity (TI) of the flow, defined as the ratio of the standard deviation of velocity fluctuations to the mean velocity from the time resolved data for each of the velocities in three directions. From these calculations we find that values for all concerned locations are in excess of 20\%, which are strongly indicative of the high turbulence levels in the flow and not originating from random noise.} To characterize the intensity of momentum and kinetic energy transfer between the two phases we calculate the turbulent kinetic energy or TKE. It is defined as half the sum of the variances (square of the standard deviation) of the velocity components \citep{Urban2017,Jedelsky2018} mathematically given by $\textrm{TKE} = 0.5 \left[\overline{(u^\prime)^2} + \overline{(v^\prime)^2} + \overline{(w^\prime)^2}\right]$, where $u^\prime$, $v^\prime$, and $w^\prime$ are the RMS velocity fluctuations of axial, radial, and tangential velocity components, respectively.
\begin{figure*}[htp!]
   \centering
   \vspace{0pt}
    \includegraphics[width=\textwidth]{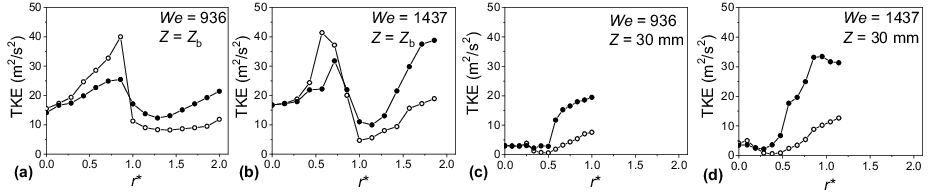}
   \vspace{-15pt}
  \caption{\label{Fig16}Variation of the TKE for the gaseous (unfilled symbols) and liquid phases (filled symbols) along the radial distance at $Z = Z_b$ and $Z =$ 30 mm for the spray with different \textit{We}. \vspace{-10pt}}
\end{figure*}

\textcolor{black}{First, we distributed all our drop size data into three different size classes, $D < 5 \mu$m, $5 \mu\textrm{m} \le D \le 25 \mu$m and $D > 25 \mu$m as shown in Figs. \hypersetup{linkcolor=black}{\ref{Fig13}} (a) and (b) for the first two size classes. The first two contain droplets of small enough diameter to faithfully follow the surrounding gas flow with Stokes number, Stk $\mathcal{O}(1)$. Regarding the size class, $5 \mu\textrm{m} \le D \le 25 \mu\textrm{m}$ only part of the drops belong to the category where Stk $\mathcal{O}(1)$. To identify this boundary, we adopt the approach given by  \hypersetup{citecolor=black}{\citep{Hinterbichler2020}} wherein the mean of the two peaks is identified as the cut off threshold. Second, we combine all the data from the size class, $D < 5 \mu\textrm{m}$ and the part relevant to our case as identified above in size class 5 $\mu\textrm{m} \le D \le 25 \mu\textrm{m}$. From this combined data we calculate the mean velocity and mean of the standard deviation $\overline{\left(u^\prime\right)^2}$, $\overline{\left(v^\prime\right)^2}$ and $\overline{\left(w^\prime\right)^2}$ (or RMS). This process is then repeated for each measurement location.}

Adopting the procedure outlined above we plot Figure \ref{Fig16} \textcolor{black}{(a)}-\textcolor{black}{(d)} to depict the radial variation of TKE of gaseous and liquid phases at $Z = Z_b$ (see Fig. \ref{Fig16} \textcolor{black}{(a)}-\textcolor{black}{(b)} and $Z =$ 30 mm (see Fig. \ref{Fig16} \textcolor{black}{(c)}-\textcolor{black}{(d)}) for the spray with different \textit{We}. 

It can be seen from the Fig. \ref{Fig16} that, for a given \textit{We}, the radial variation of TKE at $Z = Z_b$ exhibits two peaks: one is at the spray boundary and the other one is at the radial location where the droplet axial velocity changes sign from negative to positive. The lowest value of the TKE is seen at the film breakup zone due to smaller effect of inertia, for all \textit{We} in Figs. \ref{Fig16}\textcolor{black}{(a)} and \textcolor{black}{(b)} with no significant differences between the liquid and gaseous phases for the two different \textit{We} at $r^{*}$ close to the spray axis for $Z = Z_b$. Further, low value of TKE is observed at the film breakup zone despite the presence of large droplets, which is different compared to other $Z$ and previously reported observations \citep{Jedelsky2018}. From Figs. \ref{Fig16}\textcolor{black}{(c)} and \textcolor{black}{(d)} at $Z =$ 30 mm, a peak in TKE is observed at $r^{*}$ close to the spray periphery. At the spray axis, the low value of TKE is due to the absence of large, flow induced droplets. Due to larger inertia of bigger drops a notable difference in gaseous and liquid phases in the far region ($Z =$ 30 mm) of the spray is seen at higher $We =$ 1497. 
\section{Summary and conclusions}
By appropriate scaling of the spray drop size characteristics such as SMD and $U$ the spray field is divided into near region and far region of the liquid film breakup by virtue of observed self-similarity. Two axial locations are selected, one in the near-region ($Z = Z_b$) and the other is in the far-region ($Z = 30$ mm) of the liquid film breakup to compare the drop size characteristics. 
\begin{enumerate}[label=(\roman*),leftmargin=*, labelsep=1em, topsep=0pt, itemsep=-1ex,partopsep=1ex,parsep=1ex]
\item Scaling of SMD and $U$ with maximum radial Sauter Mean Diameter, SMD\textsubscript{\textit{max}} and \textit{U}\textsubscript{\textit{max}} exhibit two distinct variations and are used to demarcate near and far region of the spray which have only been approximately located in literature so far.
\item From the above, the influence of liquid film breakup on the droplet characteristics is seen to be approximately 2 to 2.5 times $L_b$ from the orifice exit. 
\item The number flux distributions in the near and far regions are differentiated by the absence of a second peak in the latter showing the memory of breakup is not retained at regions far away from the orifice exit. Observations on the maximum volume flux of the spray at a given $Z$ in the far-region reveal that it is closer to the spray boundary, possibly due to centrifuge effect.
\item From the contour plots of the resultant velocity, a high-velocity stream is seen along the spray centerline from the droplet  whose origin almost coincides with the demarcation line ($Z_{NR}$) for all \textit{We}.
\item The global PDFs integrated over radial positions for droplet size and velocity shows bi-modal distribution in the near-region and a unimodal distribution in the far region of liquid film breakup. The bimodal drop size distribution at $Z = Z_b$ is well captured by the double Gaussian distribution. On the other hand, the uni-modal drop size distribution at $Z = 30$ is accurately predicted by the gamma distribution while the log-normal distribution failed to do so. 
\item Global (radially averaged) $D_d-U_a$ correlation shows that $U_a$ at $Z = Z_b$ is independent of droplet diameter for $D_d > 25 \mu\textrm{m}$, below this $U_a$ exponentially increases with $D_d$. At $Z = 30$ mm, teaspoon effect is observed for $D_d \le 25 \mu\textrm{m}$, where $U_a$ decreases initially up to $D_d \le$ 25 $\mu\textrm{m}$ and then increases continuously with $D_d$. 
\item Based on the response of spray with the surrounding airflow the droplets are divided into small ($D_d < 5\;\mu\textrm{m}$) with $\textrm{\textit{Stk}} << 1$, medium ($5\;\mu\textrm{m} \le D_d \le 25\;\mu\textrm{m}$) with $\textrm{\textit{Stk}} = \mathcal{O}(1)$, and large ($D_d > 25\;\mu\textrm{m}$) with $\textrm{\textit{Stk}} > 1$. It was identified from the global droplet axial velocity distributions at $Z = Z_b$, that for small and medium size classes the peak of this data lies in the negative region of the droplet axial velocity whereas for the large size class it is on the positive side. 
\item Lastly, measurements on TKE reveal marked difference in gaseous and liquid phases in the far region ($Z =$ 30 mm) of the spray at higher $We =$ 1497 due to higher inertia of the large drops. In the near region, a characteristic trough is seen at the film breakup zone at where inertial effects are expected to be minimal.
\end{enumerate}
\section*{Acknowledgment}
This work is carried out with the support of National Centre for Combustion Research and Development (NCCRD), Indian Institute of Science, India. 
\section*{Declaration of competing interest}
The authors declare that they have no known competing financial interests or personal relationships that could have appeared to influence the work reported in this paper.
\section*{Data availability statement}
Experimental data presented here as plots can be provided upon request.
\appendix
\renewcommand\thefigure{A\arabic{figure}}    
\renewcommand\theequation{A\arabic{equation}}
\setcounter{table}{0}
\setcounter{figure}{0} 
\section{Choice of similarity variable $\zeta$ in demarcating near and far regions}\label{SecAppA}

\begin{figure*}[htp!]
   \centering
   \vspace{0pt}
    \includegraphics[width=\textwidth]{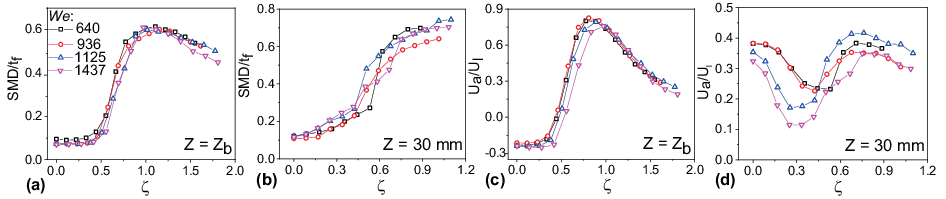}
   \vspace{-15pt}
  \caption{\label{FigA1}Comparison of SMD/$t_f$ and $U_a/U_l$ in the near and far region of liquid film breakup for different \textit{We}. (a) SMD/$t_f$ at $Z = Z_b$, (b) $U_a/U_l$ at $Z = Z_b$, (c) SMD/$t_f$ at $Z =$ 30 mm, and (d) $U_a/U_l$ at $Z =$ 30 mm.}
\end{figure*}

In the work by Dhivyaraja et al. \citep{Dhivyaraja2019} the similarity variable $\zeta$ constructed as shown in Eq. \ref{simDh} is used. There are three caveats with this selection, firstly, it is restricted to \textit{Re} $<$ 1000, in our case this value exceeds 10,000 (see Table \ref{Table1}) secondly, it is applied to the region $|\zeta| < 1/2$, which incidentally also allowed for demarcation of core of the spray radially and lastly, it is applicable for region outside the near region, $Z_{NR}$ (see Table \ref{Table2}) . Here, we check whether these restrictions can be relaxed (as anticipated) for our test conditions (see Table \ref{Table1}). To this end,in Figure \ref{FigA1} we plot the radial profiles of SMD, and $U_a$ normalized using liquid film thickness, $t_f$  and liquid sheet streamwise velocity, $U_l$ respectively against the dynamic similarity variable, $\zeta$ and given by the following, \citep{Dhivyaraja2019}.
\begin{equation}\label{simDh}
\zeta = \dfrac{r}{Z\;\textrm{tan}(\beta/2)}, \;\;\; \textrm{where,   } \beta \textrm{ is the spray cone angle}
\end{equation}								
The difference in the variation of SMD/$t_f$ and $U_a/U_l$ in the near($Z = Z_b$) and far ($Z =$ 30 mm) region of liquid film breakup are clearly seen in Fig. \ref{FigA1}. It can be seen from Fig. \ref{FigA1}\textcolor{black}{(a)} and \textcolor{black}{(b)} SMD is well scaled by $t_f$ at both $Z = Z_b$ and $Z =$ 30 mm. This collapse of data into a single line is observed when SMD is scaled with $t_f$ is observed in earlier works too \citep{Dhivyaraja2019,Vankeswaram2022} particularly in the hollow region of the spray where the number flux is high and volume flux is low as seen in Fig. \ref{Fig8}. 
While $U_a$ is scaled well by $U_l$ at $Z = Z_b$ (see Fig. \ref{FigA1}\textcolor{black}{(c)}) that is not the case for $Z =$ 30 mm (see Fig. \ref{FigA1}\textcolor{black}{(d)}). Although, a very good collapse of data (SMD/$t_f$ and $U/U_l$) in the near region of the liquid film breakup is observed, which suggest that $t_f$ and $U_l$ can be excellent scaling parameters at axial locations influenced by the liquid film breakup (near region), in the far region, only SMD data shows a good collapse. The velocity scaling is unsatisfactory as it does not show a clear differing trend like Figs. \ref{Fig3} and \ref{Fig4} which makes $r_{mvf}$ a better choice for our work.
\renewcommand\thefigure{B\arabic{figure}}  
\setcounter{figure}{0}
\section{Absence of secondary breakup at centerline of spray seen in Fig. \ref{Fig5a}}\label{SecAppB}
\begin{figure*}[htp!]
   \centering
   \vspace{0pt}
    \includegraphics[scale=0.8]{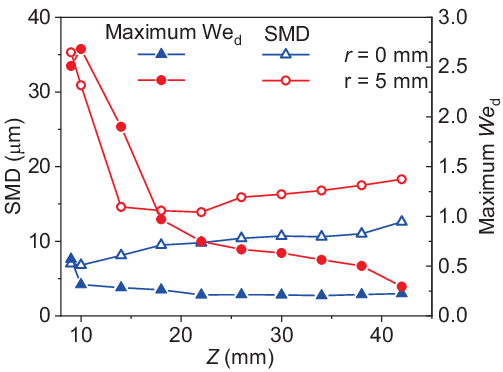}
   \vspace{-5pt}
  \caption{\label{FigB1}Variation of SMD and the corresponding droplet maximum $We_d$ at $r = 0$ and $r = 5$ mm along the axial location for $We =$ 1437.}
\end{figure*}
Secondary breakup (catastrophic being one of its types) is cascading in nature and continues until the Weber number ($We_d = \rho_gU_d^2D_d/\sigma$) of the final stable daughter droplet falls below a critical value (approximately 4-6). At this point, the breakup process terminates, and the process of droplet coalescence starts dominating \citep{Guildenbecher2009, Saha2012} which can produce drops of the range 9 - 18 $\mu$m lie. Figure \ref{FigB1} shows the variation of SMD and the droplet maximum $We_d$ along the axial direction for two radial locations, $r =$ 0 and $r =$ 5 mm at $We =$ 1437. A steady increase in SMD (approximately in the range 9 - 18 $\mu$m) at $r =$ 0 shows the absence of secondary breakup, which is well supported by the values of maximum $We_d$ ($< 1$) being relatively unchanged. At $r =$ 5 mm, a sharp decrease and then a steady increase in SMD shows that the possibility of secondary breakup initially and the same can be seen for maximum $We_d$ along $Z$. Similar behavior was observed in earlier work too along the spray centerline \citep{Saha2012} which shows that the secondary breakup is more prominent at axial locations immediate downstream of the primary breakup. Such trends are seen at other We too, which shows the absence of secondary breakup along the spray centerline.  

\renewcommand{\thetable}{C\arabic{table}}
\renewcommand\theequation{C\arabic{equation}}
\setcounter{equation}{0}
\section{Fit parameters for unimodal and bimodal Gaussian fits used in Fig. \ref{Fig6}}\label{SecAppC}
In this section we provide the fit parameters used in Fig. \ref{Fig6}. The unimodal Gaussian distribution used in Fig. \ref{Fig6}  (b),(c) and (d) is mathematically defined as,
\begin{equation}\label{GaussUni}
 \mathcal{G}(\lambda) = \mathcal{G}_0 + \dfrac{A}{w\sqrt{\pi/2}} e^{-2\left({\dfrac{\lambda-\lambda_m}{w}}\right)^2}
\end{equation}
In the above $A$, $w$, $\lambda_m$ are the fit parameters to the variables $n_N$ and $q_N$. The following table \ref{Table3a} gives values for these parameters.
\vspace{-20pt}
\begin{table}[htp!]
\small
\begin{center}
{ 
\caption{\normalsize Parameters for unimodal distribution.}\label{Table3a}
{\begin{tabular}{ccccc}
\\
\toprule
 Fig. 7 subpart & $\mathcal{G}_0$ & \hspace{0.3em} $A$ & \hspace{0.3em} $w$ & \hspace{0.3em} $\lambda_m$ \\
\midrule
 (b) & 0.15 $\pm$ 0.02 & 0.58 $\pm$ 0.04 & 0.58 $\pm$ 0.03 & 0.05\\
 (c) & 0.015 & 0.51 $\pm$ 0.03 & 0.44 $\pm$ 0.02 & 1.1 \\
 (d) & 0.086 & 0.49 $\pm$ 0.06 & 0.47 $\pm$ 0.04 & 0.92\\
\bottomrule
\end{tabular}
    }}
\end{center} 
\end{table}
\vspace{-10pt}
And the bimodal Gaussian distribution used in Figs. \ref{Fig6}(a).
\begin{equation}\label{GaussBi}
 \mathcal{G}(\lambda) = \mathcal{G}_0 + \dfrac{A}{w\sqrt{\pi/2}} e^{-2\left({\dfrac{\lambda-\lambda_m}{w}}\right)^2} + \mathcal{G}_1 + \dfrac{A_1}{w_1\sqrt{\pi/2}} e^{-2\left({\dfrac{\lambda-\lambda_{m1}}{w_1}}\right)^2}
\end{equation}
In Eq. \ref{GaussBi} the parameters $A_1$, $w_1$, $\lambda_m1$ are additional fit parameters to the variables $n_N$ while $A$, $w$, $\lambda_m$ are same as that for Eq. \ref{GaussUni}. The following table \ref{Table3b} gives values for these parameters.
\vspace{-10pt}
\begin{table}[htp!]
\small
\begin{center}
{ 
\caption{\normalsize Parameters for bimodal distribution.}\label{Table3b}
{\begin{tabular}{ccccccccc}
\\
\toprule
  Fig. 7 subpart & $\mathcal{G}_0$ & \hspace{0.3em} $A$ & \hspace{0.3em} \textit{w} & \hspace{0.3em} $\lambda_m$ & $\mathcal{G}_1$ & \hspace{0.3em} $A_1$ & \hspace{0.3em} $w_1$ & \hspace{0.3em} $\lambda_{m1}$\\
\midrule
 (a) & 0.36 $\pm$ 0.04 & 0.33 & 0.85 & 1  & -0.34 & 0.43 & 0.35 & 0.12 \\
\bottomrule
\end{tabular}
    }}
\end{center} 
\end{table}
\pagebreak
\nolinenumbers
\appto{\bibsetup}{\sloppy}
\bibliographystyle{elsarticle-harv}
\bibliography{Refs}
\clearpage
\nolinenumbers
\centerline{\textbf{HIGHLIGHTS}}
\hrulefill
\begin{center}
\begin{enumerate}[font=$\bullet$]
\item[] Near and far regions of spray are precisely demarcated using scaling analysis.
\item[] Number flux distribution shows the memory of breakup is not retained far away.
\item[]	Double Gaussian and Gamma fits predict drop size distribution in these regions.
\item[]	High velocity stream originates from the near region of the spray.
\item[]	Axial velocity for different drop size classes is different in near and far regions.
\end{enumerate}
\hspace{15pt}\hrulefill
\end{center}







\end{document}